\documentclass{iopjournal}
\usepackage{graphicx}
\usepackage{amssymb}
\usepackage{amsmath}

\usepackage{braket}
\usepackage{float}
\usepackage{ragged2e}
\usepackage{ragged2e}
\usepackage{epstopdf}
\usepackage{bm}
\usepackage{hyperref}
\usepackage[normalem]{ulem}

\begin{document}
\articletype{Paper}
\title{Exploration of Higher-order Nondegenerate Soliton solutions for Manakov model via Gram-determinant Darboux transformation}

\author{K N Santhiya$^1$  and P.~S.~Vinayagam$^{1,*}$\orcid{0000-0002-1455-7657}}

\affil{$^1$ Department of Physics, PSG College of Arts and Science (Autonomous), Avinashi Road, Civil Aerodrome Post, Coimbatore 641 014, Tamil Nadu, India}

\affil{$^*$Author to whom any correspondence should be addressed.}

\email{psvinayagam11@gmail.com}


\begin{abstract}
Higher-order nondegenerate soliton solutions of the integrable Coupled Nonlinear Schr\"odinger system, known as the Manakov system, are obtained using the Darboux transformation associated with the Gram-determinant approach. Starting from the trivial seed solution, the explicit first and second-order soliton solutions are constructed through the direct iterative Darboux procedure. Owing to the rapid increase in algebraic complexity with successive iterations, the higher-order solutions are reformulated through a compact Gram-determinant representation, thereby establishing a unified and systematic framework for their construction. Using this approach, explicit third- and fourth-order solutions are derived, and the general $N$th-order case is presented. The resulting structures exhibit asymmetric localized profiles with multiple intensity peaks arising from distinct spectral parameters. The collision dynamics of higher-order nondegenerate solitons reveal features such as asymmetric energy redistribution and complex interaction behaviour. The present formulation provides a systematic and explicit construction of higher-order nondegenerate solitons, offering new insights into multi-component nonlinear wave dynamics. 
\end{abstract}
\section{Introduction}
The study of nonlinear partial differential equations (NPDEs) plays a fundamental role in understanding complex wave propagation phenomena in nature. The interplay between nonlinearity and dispersion gives rise to a wide variety of localized structures such as solitons, breathers, and other nonlinear wave patterns \cite{ref1,ref2,ref3,ref4}. NPDEs also provide an important theoretical framework for investigating wave localization \cite{ref5,ref6}, pattern formation \cite{ref7}, modulation instability (MI) \cite{ref8}, and energy exchange mechanisms in interacting nonlinear systems \cite{ref9,ref10}. Among the various nonlinear evolution equations, the nonlinear Schr\"odinger (NLS) equation is one of the most important models describing the slowly varying envelope of wave packets in weakly nonlinear dispersive media \cite{ref11,ref12}. The balance between dispersion and nonlinearity leads to the formation of localized wave structures known as solitons \cite{ref13,ref14,ref15}. These solitons preserve their shape during propagation and exhibit particle-like interaction behaviour. Owing to these remarkable properties, the NLS equation has found wide applications in nonlinear optics \cite{ref16,ref17}, plasma physics \cite{ref18,ref19,ref20}, and fluid dynamics \cite{ref21,ref22,ref23}.
The integrability of nonlinear wave equations through the inverse scattering transform (IST) was pioneered by Zakharov and Shabat in their seminal work on self-focusing and nonlinear wave modulation \cite{ref24}. Their formulation established the mathematical foundation for several integrable nonlinear evolution equations, including coupled nonlinear Schr\"odinger CNLS systems such as the Manakov model. Extending the scalar NLS equation to CNLS systems provides a natural framework for studying multi-component wave interactions. The Manakov system serves as an important theoretical model for nonlinear optical fibres in which two orthogonally polarized modes interact through nonlinear effects \cite{ref25,ref26}. One of the most significant properties of the Manakov system is its integrability, which allows the existence of vector soliton solutions \cite{ref27}. Unlike scalar solitons, vector solitons involve multiple coupled wave components that propagate coherently while preserving their localized nature. During interaction, these vector solitons exhibit several intriguing phenomena such as energy-sharing interactions \cite{ref28,ref29}, shape-changing collisions \cite{ref30,ref31}, and the formation of bound states \cite{ref32,ref33}. These remarkable features have attracted considerable attention in nonlinear optics because of their potential applications in optical switching \cite{ref34,ref35}, signal control, and information processing in fibre communication systems \cite{ref36,ref37}.

The Manakov system admits two distinct classes of vector solitons, namely degenerate and nondegenerate solitons. This classification is determined by the spectral parameters associated with the soliton components. In the degenerate case, both components are associated with identical eigenvalues, $\lambda_1=\lambda_2$, resulting in symmetric intensity distributions across the components. Such solutions have been extensively investigated, and their collision dynamics are relatively well understood \cite{ref38}. In contrast, nondegenerate solitons arise when the corresponding eigenvalues are distinct, i.e., $\lambda_1 \neq \lambda_2$ \cite{ref39,ref40}. The presence of distinct spectral parameters introduces additional degrees of freedom, leading to richer localized wave structures and more intricate nonlinear dynamics. Consequently, nondegenerate solitons can exhibit complex localized profiles such as bound states \cite{ref41}, flat-top structures \cite{ref42}, and composite localized wave patterns \cite{ref43}. These additional dynamical features significantly enrich the solution space of CNLS systems and provide deeper insight into multi-component nonlinear wave interactions.

The study of vector solitons in CNLS systems began with the introduction of the integrable two-component NLS equation through the IST \cite{ref44}. Subsequently, the general two-soliton solution and its collision dynamics were obtained using the Hirota bilinear method \cite{ref45}. Further investigations employed the Hirota approach together with asymptotic analysis to study multisoliton solutions and energy-sharing interactions \cite{ref46}. It is worth mentioning several physically motivated studies~\cite{ref47,ref48} suggested generalized multi-component NLS-type integrable systems on a spatially continuous multi-fibre support, where additional linear couplings between fibers responsible for the coherent intercomponent interaction effects have been incorporated. Higher-order solutions were later investigated using the Darboux transformation (DT) \cite{ref49}. The existence of nondegenerate solitons in the Manakov system was reported through the Hirota bilinear method \cite{ref50}, while nondegenerate bound-state solitons in integrable coupled NLS equation were studied using the DT \cite{ref51}. Further studies examined higher-order nondegenerate solitons and their collision dynamics using the Hirota bilinear method with Gram-determinant representations \cite{ref52,ref53}. More recently, these investigations have been extended to fractional and multi-component integrable systems using approaches such as IST, DT, and Kadomtsev--Petviashvili (KP) hierarchy reduction methods \cite{ref54,ref55}.

Although several analytical techniques, including the DT, Hirota bilinear method, Wronskian formulations, and Grammian approaches, have been successfully employed to investigate the Manakov system, most existing studies are primarily restricted to fundamental or lower-order soliton structures. In many cases, higher-order solutions are presented only in formal generalized forms without explicitly analysing their detailed analytical structure and interaction dynamics. These limitations motivate the need for a systematic and computationally efficient framework capable of constructing explicit higher-order nondegenerate soliton solutions. In the present work, we adapt the Gram-determinant representation within the DT framework to systematically derive higher-order nondegenerate soliton solutions of arbitrary order. The proposed formulation significantly reduces the algebraic complexity associated with direct iterative DT calculations while enabling the explicit construction of higher-order localized structures beyond the commonly studied lower-order cases. Using this framework, we further investigate the collision dynamics of higher-order nondegenerate solitons and demonstrate that the presence of multiple independent spectral parameters and amplitude coefficients gives rise to rich nonlinear behaviours such as symmetric and asymmetric localized wave profiles, velocity-locked propagation, elastic interactions, and inelastic collision dynamics. The investigation of these nonlinear wave structures provides deeper insight into complex wave propagation phenomena in multi-component integrable systems.

In this work, we develop a unified and systematic framework based on the DT combined with a Gram-determinant representation to construct higher-order nondegenerate soliton solutions in a compact and computationally tractable form. Unlike conventional iterative DT approaches, the present formulation enables explicit analytical construction of higher-order nondegenerate solitons with significantly reduced algebraic complexity. Furthermore, we investigate the interaction dynamics of higher-order nondegenerate solitons and reveal several rich nonlinear behaviours, including asymmetric energy redistribution, evolution of complex localized profiles, and velocity-locked bound states arising from distinct spectral parameters. These results demonstrate that higher-order nondegenerate solitons possess substantially richer dynamical properties compared to their degenerate counterparts. The findings presented in this work provide new insight into the structure and interaction of multi-component nonlinear waves and establish a systematic framework for exploring higher-order nonlinear phenomena in integrable coupled systems.

The rest of the paper is organized as follows. Section 2 introduces the integrable CNLS system together with its associated Lax pair representation. Section 3 presents the construction of first- and second-order nondegenerate soliton solutions using the DT with a trivial seed solution. A Gram-determinant representation of the $N$-fold DT is then formulated to generate higher-order nondegenerate soliton solutions systematically. Sections 4 and 5 analyse the behaviour and collision dynamics of higher-order nondegenerate solitons up to fifth order. Finally, Section 6 summarizes the main findings of the present work.

\section{Model Equation and Its Lax Pair Representation}
\vspace{\baselineskip}
We start by introducing the integrable Coupled Nonlinear Schrödinger (CNLS) system that serves as the mathematical model throughout this work. This system describes the propagation of two interacting complex wave envelopes under the combined influence of second-order dispersion and nonlinear self- and cross-phase modulation. The governing equations take the form
\begin{subequations}
\begin{align}
i\,\frac{\partial q_1(x,t)}{\partial t}
+ \frac{\partial^2 q_1(x,t)}{\partial x^2}
+ 2\left(\lvert q_1(x,t) \rvert^2 + \lvert q_2(x,t) \rvert^2\right) q_1(x,t) &= 0, \label{eq1a} \\
i\,\frac{\partial q_2(x,t)}{\partial t}
+ \frac{\partial^2 q_2(x,t)}{\partial x^2}
+ 2\left(\lvert q_1(x,t) \rvert^2 + \lvert q_2(x,t) \rvert^2\right) q_2(x,t) &= 0. \label{eq1b} 
\end{align}
\end{subequations}
Here $q_j(x,t)$ $(j=1,2)$ denote the two coupled wave components, $x$ is the spatial coordinate, and $t$ represents the evolution variable. The nonlinear terms arise from equal contributions from self-phase and cross-phase modulation, a balance that is crucial to the system's integrability. Eqs.~\eqref{eq1a}--\eqref{eq1b}, referred to as the Manakov equations \cite{ref25}, describe the integrable CNLS system.
A key feature of the Manakov system is its complete integrability, which can be established through the existence of a Lax pair. The nonlinear system can be recast as the compatibility condition of the following overdetermined linear system:
%\begin{subequations}
\begin{align}
\boldsymbol{\Phi}_x &= \mathbf{U}\,\boldsymbol{\Phi}\label{lax_spatial}, \\
\boldsymbol{\Phi}_t &= \mathbf{V}\,\boldsymbol{\Phi}\label{lax_temporal},
\end{align}
%\end{subequations}
where $\boldsymbol{\Phi}=(\phi_1,\phi_2,\phi_3)^{T}$ is the three-component auxiliary function and $\lambda \in \mathbb{C}$ denotes the spectral parameter. The spatial evolution matrix $\mathbf{U}$ is given by
\begin{equation}
\mathbf{U} =
\begin{bmatrix}
-\dfrac{2i}{3}\lambda & q_1 & q_2 \\
- q_1^{*} & \dfrac{i}{3}\lambda & 0 \\
- q_2^{*} & 0 & \dfrac{i}{3}\lambda
\end{bmatrix},
\end{equation}
and the corresponding temporal evolution matrix $\mathbf{V}$ takes the form

\begin{equation}
\mathbf{V} =
\begin{bmatrix}
-\dfrac{2i}{3}\lambda^{2} + i\left(\lvert q_1\rvert^{2}+\lvert q_2\rvert^{2}\right)
& 2\lambda q_1 + i q_{1x}
& 2\lambda q_2 + i q_{2x} \\[6pt]

-2\lambda q_1^{*} + i q_{1x}^{*}
& \dfrac{i}{3}\lambda^{2} - i \lvert q_1\rvert^{2}
& - i q_1^{*} q_2 \\[6pt]

-2\lambda q_2^{*} + i q_{2x}^{*}
& - i q_2^{*} q_1
& \dfrac{i}{3}\lambda^{2} - i \lvert q_2\rvert^{2}
\end{bmatrix}.
\end{equation}
The compatibility (zero-curvature) condition
\begin{equation}
\mathbf{U}_t-\mathbf{V}_x+[\mathbf{U},\mathbf{V}]=\mathbf{0},
\end{equation}
reproduces exactly the coupled Eqs.~\eqref{eq1a}--\eqref{eq1b}. The existence of the above Lax representation establishes the complete integrability of the Manakov system and ensures the availability of a rich class of exact analytical solutions. The integrable structure plays a fundamental role in determining the qualitative characteristics of the resulting nonlinear wave solutions. In particular, the spectral parameter $\lambda$ serves as the key quantity governing the construction and properties of soliton states through the associated linear spectral problem. Different choices of spectral parameters generate localized vector solitons with distinct amplitude distributions and localization characteristics in the two wave components. When the spectral parameters are chosen as distinct complex values, the corresponding solutions represent nondegenerate vector solitons that may form asymmetric bound-state configurations and exhibit a variety of interaction behaviours. In the next section, we employ the Darboux transformation, one of the most powerful analytical techniques for constructing exact solutions of integrable nonlinear evolution equations \cite{ref56,ref57,ref58}. Starting from a simple seed solution and utilizing the above Lax pair, the Darboux transformation enables the recursive generation of higher-order soliton solutions while preserving the integrable structure of the system. As the iterative procedure becomes increasingly involved at higher orders, the resulting solutions are subsequently reformulated in terms of a compact Gram-determinant representation, providing a systematic framework for constructing arbitrary-order nondegenerate vector-soliton solutions.
%%%%%%%%%%%%%%%%%%%%%%%%%%%%%%%%%%%%%%%%5
\section{Explicit Construction of First- and Second-Order Soliton Solutions via Darboux Transformation}
\vspace{\baselineskip}
The procedure begins with the trivial (vacuum) seed solution,
\begin{equation}
q_1^{[0]}(x,t)=0, \qquad q_2^{[0]}(x,t)=0,
\end{equation}
substituting the zero seed into the Lax pair, the spatial matrix $\mathbf{U}$ reduces to

\begin{equation}
\mathbf{U}_0=
\begin{pmatrix}
-\dfrac{2i}{3}\lambda & 0 & 0\\
0 & \dfrac{i}{3}\lambda & 0\\
0 & 0 & \dfrac{i}{3}\lambda
\end{pmatrix},
\end{equation}
similarly, the temporal matrix $\mathbf{V}$ reduces to
\begin{equation}
\mathbf{V}_0=
\begin{pmatrix}
-\dfrac{2i}{3}\lambda^2 & 0 & 0\\
0 & \dfrac{i}{3}\lambda^2 & 0\\
0 & 0 & \dfrac{i}{3}\lambda^2
\end{pmatrix},
\end{equation}
thus, the reduced Lax pair for the zero seed becomes
\begin{align}
\boldsymbol{\Phi}_x &= \mathbf{U_0}\boldsymbol{\Phi}, \label{lax_x}\\
\boldsymbol{\Phi}_t &= \mathbf{V_0}\boldsymbol{\Phi}. \label{lax_t}
\end{align}
Let
\begin{equation}
\boldsymbol{\Phi}=
\begin{pmatrix}
\phi_1\\
\phi_2\\
\phi_3
\end{pmatrix}.
\end{equation}
From Eq.~\eqref{lax_x}, the system of linear equations for the components of $\Phi$ is obtained. 
Integrating the $x$-part and substituting the resulting expressions into the $t$-part of Eq.~\eqref{lax_t}, the auxiliary function corresponding to the zero seed solution is obtained as
\begin{equation}
\boldsymbol{\phi}(x,t,\lambda)=
\begin{pmatrix}
e^{-\frac{2i}{3}(\lambda x+\lambda^2 t)}\\
c_1\,e^{\frac{i}{3}(\lambda x+\lambda^2 t)}\\
c_2\,e^{\frac{i}{3}(\lambda x+\lambda^2 t)}
\end{pmatrix}.
\end{equation}
To generate localized soliton solutions, we fix the spectral parameter as
\begin{equation}
\lambda_1=\epsilon_1+i\eta_1, \qquad \eta_1>0,
\end{equation}
where $\epsilon_1$ and $\eta_1$ are real constants.
Substituting $\lambda=\lambda_1$ into the auxiliary function, we define
\begin{equation}
\boldsymbol{\phi}_1=\boldsymbol{\phi}(x,t,\lambda_1)=
\begin{pmatrix}
\phi_{1,1}\\
\phi_{1,2}\\
\phi_{1,3}
\end{pmatrix},
\end{equation}
with
\[
\phi_{1,1}=e^{-\frac{2i}{3}(\lambda_1 x+\lambda_1^{2} t)},
\qquad
\phi_{1,2}=c_{1}\,e^{\frac{i}{3}(\lambda_1 x+\lambda_1^{2} t)},
\qquad
\phi_{1,3}=c_{2}\,e^{\frac{i}{3}(\lambda_1 x+\lambda_1^{2} t)},
\]
and the Hermitian conjugate of $\boldsymbol{\phi}_1$ is given by
$$
\boldsymbol{\phi}_1^{\dagger} =
\begin{pmatrix}
\phi_{1,1}^{*} & \phi_{1,2}^{*} & \phi_{1,3}^{*}
\end{pmatrix},
$$
where
\[
\phi_{1,1}^{*} =
e^{\frac{2i}{3}(\lambda_1^{*} x+(\lambda_1^{*})^{2} t)},
\qquad
\phi_{1,2}^{*} =
c_1^{*}\,e^{-\frac{i}{3}(\lambda_1^{*} x+(\lambda_1^{*})^{2} t)},
\qquad
\phi_{1,3}^{*} =
c_2^{*}\,e^{-\frac{i}{3}(\lambda_1^{*} x+(\lambda_1^{*})^{2} t)}.
\]
The scalar quantity appearing in the denominator of the projection operator is
$$
\boldsymbol{\phi}_1^{\dagger}\boldsymbol{\phi}_1
=
|\phi_{1,1}|^2+|\phi_{1,2}|^2+|\phi_{1,3}|^2,
$$
substituting the expressions of $\phi_{1,1}, \phi_{1,2}, \phi_{1,3}$, we obtain
$$
\boldsymbol{\phi}_1^{\dagger}\boldsymbol{\phi}_1 =
e^{-\frac{4}{3}\eta_1(x+2\epsilon_1 t)}
+
|c_1|^2 e^{\frac{4}{3}\eta_1(x+2\epsilon_1 t)}
+
|c_2|^2 e^{\frac{4}{3}\eta_1(x+2\epsilon_1 t)},
$$
using $\boldsymbol{\phi}_1$ and its Hermitian conjugate, we define the rank-one projection operator
$$
P_1=
\frac{\boldsymbol{\phi}_1\,\boldsymbol{\phi}_1^{\dagger}}
{\boldsymbol{\phi}_1^{\dagger}\boldsymbol{\phi}_1},
$$
explicitly, the matrix elements of $P_1$ are
$$
(P_1)_{ij}=
\frac{\phi_{1,i}\phi_{1,j}^{*}}
{|\phi_{1,1}|^2+|\phi_{1,2}|^2+|\phi_{1,3}|^2},
\qquad i,j=1,2,3.
$$
in particular, the elements relevant for the field variables are
\[
(P_1)_{12} =
\frac{\phi_{1,1}\phi_{1,2}^{*}}
{|\phi_{1,1}|^2+|\phi_{1,2}|^2+|\phi_{1,3}|^2},
\qquad
(P_1)_{13} =
\frac{\phi_{1,1}\phi_{1,3}^{*}}
{|\phi_{1,1}|^2+|\phi_{1,2}|^2+|\phi_{1,3}|^2}.
\]
The one-fold Darboux transformation matrix is defined as
\begin{equation}
T_1(\lambda) =
I-\frac{\lambda_1-\lambda_1^{*}}{\lambda-\lambda_1^{*}}\,P_1
\end{equation}
and the Darboux-transformed Lax matrix is given by
\begin{equation}
U^{[1]}=T_{1,x}T_1^{-1}+T_1U_0T_1^{-1}.
\end{equation}
By definition, the transformed Lax matrix must retain the Manakov form
$$
U^{[1]}=
\begin{pmatrix}
-\dfrac{2i}{3}\lambda & q_1^{[1]} & q_2^{[1]}\\
- q_1^{[1]*} & \dfrac{i}{3}\lambda & 0\\
- q_2^{[1]*} & 0 & \dfrac{i}{3}\lambda
\end{pmatrix},
$$
evaluating the off-diagonal entries of $U^{[1]}$ and collecting the $(1,2)$ and $(1,3)$ components, we obtain
\begin{align}
q_1^{[1]} &= (U^{[1]})_{12}
=
2i(\lambda_1^{*}-\lambda_1)(P_1)_{12}, \\
q_2^{[1]} &= (U^{[1]})_{13}
=
2i(\lambda_1^{*}-\lambda_1)(P_1)_{13},
\end{align}
using the explicit form of the projector elements,
$$
(P_1)_{12}=\frac{\phi_{1,1}\phi_{1,2}^{*}}
{|\phi_{1,1}|^{2}+|\phi_{1,2}|^{2}+|\phi_{1,3}|^{2}},
\qquad
(P_1)_{13}=\frac{\phi_{1,1}\phi_{1,3}^{*}}
{|\phi_{1,1}|^{2}+|\phi_{1,2}|^{2}+|\phi_{1,3}|^{2}},
$$
the first-order field variables are finally obtained as
\begin{subequations}
    \begin{align}
q_1^{[1]}&=
2i(\lambda_1^{*}-\lambda_1)
\frac{\phi_{1,1}\phi_{1,2}^{*}}
{|\phi_{1,1}|^{2}+|\phi_{1,2}|^{2}+|\phi_{1,3}|^{2}},\\
q_2^{[1]}&=
2i(\lambda_1^{*}-\lambda_1)
\frac{\phi_{1,1}\phi_{1,3}^{*}}
{|\phi_{1,1}|^{2}+|\phi_{1,2}|^{2}+|\phi_{1,3}|^{2}},
\end{align}
\end{subequations}
substituting the corrected eigenfunction components,
$$
\phi_{1,1} =
e^{-\frac{2i}{3}(\lambda_1 x+\lambda_1^2 t)},
\qquad
\phi_{1,2}^{*} =
c_1^{*}e^{-\frac{i}{3}(\lambda_1^{*} x+(\lambda_1^{*})^{2} t)},
\qquad
\phi_{1,3}^{*} =
c_2^{*}e^{-\frac{i}{3}(\lambda_1^{*} x+(\lambda_1^{*})^{2} t)},
$$
the first-order fields become

\begin{subequations}
\begin{align}
q_1^{[1]} &=
2i(\lambda_1^{*}-\lambda_1)
\frac{c_1^{*}
e^{-\frac{i}{3}(2\lambda_1+\lambda_1^{*})x
-\frac{i}{3}(2\lambda_1^2+(\lambda_1^{*})^2)t}}
{D_1},\label{q1_1}
\\
q_2^{[1]} &=
2i(\lambda_1^{*}-\lambda_1)
\frac{c_2^{*}
e^{-\frac{i}{3}(2\lambda_1+\lambda_1^{*})x
-\frac{i}{3}(2\lambda_1^2+(\lambda_1^{*})^2)t}}
{D_1},\label{q2_1}
\end{align}
\end{subequations}
where
$$
D_1=
|\phi_{1,1}|^2+|\phi_{1,2}|^2+|\phi_{1,3}|^2.
$$
Using $\lambda_1=\epsilon_1+i\eta_1$, we have
$$
|\phi_{1,1}|^2 =
e^{-\frac{4}{3}\eta_1(x+2\epsilon_1 t)},
\qquad
|\phi_{1,2}|^2 =
|c_1|^2 e^{\frac{4}{3}\eta_1(x+2\epsilon_1 t)},
\qquad
|\phi_{1,3}|^2 =
|c_2|^2 e^{\frac{4}{3}\eta_1(x+2\epsilon_1 t)},
$$
hence,
$$
D_1=
e^{-\frac{4}{3}\eta_1\xi}
+
\left(|c_1|^2+|c_2|^2\right)
e^{\frac{4}{3}\eta_1\xi},
\qquad
\xi=x+2\epsilon_1 t,
$$
factoring out $e^{-\frac{4}{3}\eta_1\xi}$ from the denominator, we obtain
\begin{equation}
D_1=
e^{-\frac{4}{3}\eta_1\xi}
\left[
1+\left(|c_1|^2+|c_2|^2\right)
e^{\frac{8}{3}\eta_1\xi}
\right],
\end{equation}
substituting the $D_1$  into Eqs.~\eqref{q1_1} and \eqref{q2_1}, we obtain
\begin{subequations}
    \begin{align}
q_1^{[1]} &=
2i(\lambda_1^{*}-\lambda_1)c_1^{*}
\frac{e^{\frac{4}{3}\eta_1\xi}}
{1+\left(|c_1|^2+|c_2|^2\right)e^{\frac{8}{3}\eta_1\xi}}
e^{-\frac{2i}{3}[\epsilon_1 x+2(\epsilon_1^2-\eta_1^2)t]},\label{q1}\\
q_2^{[1]} &=
2i(\lambda_1^{*}-\lambda_1)c_2^{*}
\frac{e^{\frac{4}{3}\eta_1\xi}}
{1+\left(|c_1|^2+|c_2|^2\right)e^{\frac{8}{3}\eta_1\xi}}
e^{-\frac{2i}{3}[\epsilon_1 x+2(\epsilon_1^2-\eta_1^2)t]},\label{q2}
\end{align}
\end{subequations}
where $\xi = x+2\epsilon_1 t$.
By rewriting the exponential factor in Eqs.~\eqref{q1} and \eqref{q2} in terms of the hyperbolic secant function, the first-order solution can be written as,
\begin{subequations}
\begin{align}
q_1^{[1]}(x,t) &=
2\eta_1 \alpha_1
\operatorname{sech}
\!\left(
\frac{4}{3}\eta_1 (x + 2\epsilon_1 t + \delta_0)
\right)
e^{-\,\frac{2i}{3}
\left[
\epsilon_1 x + 2(\epsilon_1^2 - \eta_1^2)t
\right]}, \label{firstorder_1}
\\
q_2^{[1]}(x,t) &=
2\eta_1 \alpha_2
\operatorname{sech}
\!\left(
\frac{4}{3}\eta_1 (x + 2\epsilon_1 t + \delta_0)
\right)
e^{-\,\frac{2i}{3}
\left[
\epsilon_1 x + 2(\epsilon_1^2 - \eta_1^2)t
\right]},\label{firstorder_2}
\end{align}
\end{subequations}
where
\[
\alpha_1 = \frac{c_1^*}{\sqrt{K}},
\qquad
\alpha_2 = \frac{c_2^*}{\sqrt{K}},
\qquad
\delta_0 = -\frac{3}{8\eta_1}\ln K.
\]
Eqs.~\eqref{firstorder_1} and \eqref{firstorder_2} represent a degenerate vector soliton of first order in the Manakov system.
To proceed further, we introduce a second spectral parameter of the form
\begin{equation}
\lambda_2=\epsilon_1+i\eta_2, \qquad \eta_2\neq\eta_1,
\end{equation}
where the real part is chosen to be the same as that of $\lambda_1$. This choice ensures that the corresponding soliton components propagate with the same velocity, which is essential for the formation of bound-state solitons. Now we choose the auxiliary function associated with $\lambda=\lambda_2$ obtained from the Lax pair as
$$
\boldsymbol{\phi}_2=
\boldsymbol{\phi}(x,t,\lambda_2)=
\begin{pmatrix}
e^{-\frac{2i}{3}(\lambda_2 x+\lambda_2^2 t)}\\
c_{21}\,e^{\frac{i}{3}(\lambda_2 x+\lambda_2^2 t)}\\
c_{22}\,e^{\frac{i}{3}(\lambda_2 x+\lambda_2^2 t)}
\end{pmatrix}.
$$
The eigenfunction $\boldsymbol{\phi}_2$ is now mapped through the one-fold Darboux transformation constructed from $\lambda_1$ and the transformed eigenfunction is defined as
\begin{equation}
\widetilde{\boldsymbol{\phi}}_2 = T_1(\lambda_2)\boldsymbol{\phi}_2,
\end{equation}
where
$$
T_1(\lambda)=
I-\frac{\lambda_1-\lambda_1^{*}}{\lambda-\lambda_1^{*}}\,P_1,
$$
substituting $\lambda=\lambda_2$, we obtain
\begin{equation}
\widetilde{\boldsymbol{\phi}}_2 =
\left(
I-\frac{\lambda_1-\lambda_1^{*}}{\lambda_2-\lambda_1^{*}}\,P_1
\right)\boldsymbol{\phi}_2,
\end{equation}
using the definition of the projection operator $P_1$, the transformed eigenfunction $\widetilde{\boldsymbol{\phi}}_2$ can be written as
$$
\widetilde{\boldsymbol{\phi}}_2 =
\boldsymbol{\phi}_2
-
\frac{\lambda_1-\lambda_1^{*}}{\lambda_2-\lambda_1^{*}}
\frac{\boldsymbol{\phi}_1
\left(\boldsymbol{\phi}_1^{\dagger}\boldsymbol{\phi}_2\right)}
{\boldsymbol{\phi}_1^{\dagger}\boldsymbol{\phi}_1},
$$
from which the individual components are obtained explicitly as
\begin{subequations}
\begin{align}
\widetilde{\phi}_{2,1} &=
\phi_{2,1}
-
\frac{\lambda_1-\lambda_1^{*}}{\lambda_2-\lambda_1^{*}}
\frac{
\phi_{1,1}
\left(
\phi_{1,1}^{*}\phi_{2,1}
+
\phi_{1,2}^{*}\phi_{2,2}
+
\phi_{1,3}^{*}\phi_{2,3}
\right)}
{|\phi_{1,1}|^{2}+|\phi_{1,2}|^{2}+|\phi_{1,3}|^{2}},
\label{65a}
\\
\widetilde{\phi}_{2,2} &=
\phi_{2,2}
-
\frac{\lambda_1-\lambda_1^{*}}{\lambda_2-\lambda_1^{*}}
\frac{
\phi_{1,2}
\left(
\phi_{1,1}^{*}\phi_{2,1}
+
\phi_{1,2}^{*}\phi_{2,2}
+
\phi_{1,3}^{*}\phi_{2,3}
\right)}
{|\phi_{1,1}|^{2}+|\phi_{1,2}|^{2}+|\phi_{1,3}|^{2}},
\label{65b}
\\
\widetilde{\phi}_{2,3} &=
\phi_{2,3}
-
\frac{\lambda_1-\lambda_1^{*}}{\lambda_2-\lambda_1^{*}}
\frac{
\phi_{1,3}
\left(
\phi_{1,1}^{*}\phi_{2,1}
+
\phi_{1,2}^{*}\phi_{2,2}
+
\phi_{1,3}^{*}\phi_{2,3}
\right)}
{|\phi_{1,1}|^{2}+|\phi_{1,2}|^{2}+|\phi_{1,3}|^{2}}.
\label{65c}
\end{align}
\end{subequations}
Using the transformed eigenfunction $\widetilde{\boldsymbol{\phi}}_2$, the second rank-one projection operator is defined as

$$
P_2=
\frac{\widetilde{\boldsymbol{\phi}}_2\, \widetilde{\boldsymbol{\phi}}_2^{\dagger}}
{\widetilde{\boldsymbol{\phi}}_2^{\dagger} \widetilde{\boldsymbol{\phi}}_2},
$$
the denominator of the projection operator is given explicitly by
$$
\widetilde{\boldsymbol{\phi}}_2^{\dagger} \widetilde{\boldsymbol{\phi}}_2
=
|\widetilde{\phi}_{2,1}|^2
+
|\widetilde{\phi}_{2,2}|^2
+
|\widetilde{\phi}_{2,3}|^2,
$$
where each component $\widetilde{\phi}_{2,i}$ is defined in Eqs.~\eqref{65a}–\eqref{65c}. Using $\lambda_j=\epsilon_1+i\eta_j$ $(j=1,2)$ and $\xi=x+2\epsilon_1 t$, the exponential structure of the denominator can be expressed in terms of $\exp\!\left(\frac{4}{3}\eta_j \xi\right)$, and after simplification the denominator takes the general form
\begin{equation}
\mathcal{D}(\xi)
=
1
+
A_1 e^{\frac{8}{3}\eta_1\xi}
+
A_2 e^{\frac{8}{3}\eta_2\xi}
+
A_{12} e^{\frac{8}{3}(\eta_1+\eta_2)\xi},
\end{equation}
where the constants $A_1$, $A_2$, and $A_{12}$ depend on $c_1$, $c_2$, $c_{21}$, $c_{22}$ and the spectral parameters, and the relevant projector elements are written as
$$
(P_2)_{12}
=
\frac{\widetilde{\phi}_{2,1}\widetilde{\phi}_{2,2}^{*}}
{\mathcal{D}(\xi)},
\qquad
(P_2)_{13}
=
\frac{\widetilde{\phi}_{2,1}\widetilde{\phi}_{2,3}^{*}}
{\mathcal{D}(\xi)},
$$
accordingly, the second-order iterated fields are obtained as
\begin{equation}
q_1^{[2]}
=
q_1^{[1]}
+
2i(\lambda_2^{*}-\lambda_2)(P_2)_{12},
\qquad
q_2^{[2]}
=
q_2^{[1]}
+
2i(\lambda_2^{*}-\lambda_2)(P_2)_{13}.
\end{equation}
Using
\[
\lambda_2^{*}-\lambda_2=-2i\eta_2,
\]
we obtain
\[
2i(\lambda_2^{*}-\lambda_2)=4\eta_2,
\]
hence,
\begin{subequations}
\begin{align}
q_1^{[2]}(x,t)
&=
q_1^{[1]}(x,t)
+
4\eta_2
\frac{\widetilde{\phi}_{2,1}\widetilde{\phi}_{2,2}^{*}}
{\widetilde{\boldsymbol{\phi}}_2^{\dagger}\widetilde{\boldsymbol{\phi}}_2}
\exp\!\left[
-\frac{2i}{3}
\left(
\epsilon_1 x
+
2(\epsilon_1^2-\eta_2^2)t
\right)
\right],\label{q1_second order}
\\ 
q_2^{[2]}(x,t)
&=
q_2^{[1]}(x,t)
+
4\eta_2
\frac{\widetilde{\phi}_{2,1}\widetilde{\phi}_{2,3}^{*}}
{\widetilde{\boldsymbol{\phi}}_2^{\dagger}\widetilde{\boldsymbol{\phi}}_2}
\exp\!\left[
-\frac{2i}{3}
\left(
\epsilon_1 x
+
2(\epsilon_1^2-\eta_2^2)t
\right)
\right].\label{q2_second order}
\end{align}
\end{subequations}
The above Eqs.~\eqref{q1_second order} and \eqref{q2_second order} represent the second-order nondegenerate soliton solution of the Manakov system.

As the order of the Darboux transformation increases, the direct iterative construction becomes algebraically tedious due to the repeated appearance
of projection operators and transformed eigenfunctions at each iteration step. In particular, every successive transformation introduces additional inner products between the eigenfunctions associated with different spectral parameters. These interaction terms become increasingly complicated when higher-order nondegenerate soliton structures are considered. To overcome this difficulty, the iterative Darboux method can be reorganized into a systematic Gram-determinant representation. In this formulation, the repeated inner products generated through successive Darboux iterations are systematically collected into a structured Gram matrix. The determinant structure naturally encodes the nonlinear interaction among multiple eigenfunctions while preserving the integrable structure of the
Manakov system. As a result, the higher-order soliton solutions can be
constructed in a compact and computationally tractable form without explicitly performing all intermediate Darboux iterations. The bordered determinant representation further provides a convenient way to
express the field components of the N-fold transformed solution. Once the
Gram matrix structure is established, higher-order nondegenerate solitons of
arbitrary order can be generated systematically by enlarging the matrix
dimension, making the approach particularly suitable for studying complex 
multi-soliton interaction dynamics.

\subsection{Gram-Determinant Representation of the $N$-Fold Darboux Transformation}

To construct higher-order nondegenerate soliton solutions of the Manakov
system, we employ the Gram-determinant representation associated with the
N-fold Darboux transformation. Earlier studies developed determinant
formulations for multisoliton solutions and related Grammian structures in
integrable nonlinear systems \cite{ref51,ref59,ref60}. Motivated by these established
determinant representations, the iterative Darboux transformation can be
reorganized into Gram-determinant form. In this formulation, the higher-order nondegenerate multi-soliton structure is encoded in a finite-dimensional matrix whose entries describe pairwise interactions between eigenfunctions of the Lax pair. Once this matrix structure is fixed, solutions of arbitrary order can be generated systematically by enlarging its dimension, without repeating the full iterative Darboux procedure.
In the framework of the Darboux transformation, repeated application of the  one-fold transformation leads to a sequence of projection operators constructed from the eigenfunctions of the Lax pair. When the transformation is iterated $N$ times, the resulting expressions for the transformed fields involve combinations of inner products between the eigenfunctions. These inner products naturally form a Gram-type matrix structure. By organizing these terms systematically, the $N$-fold Darboux transformation can be rewritten in terms of Gram-determinants, which provides a compact representation of the higher-order solutions. This determinant formulation avoids the rapidly growing algebraic complexity associated with direct iterative calculations and offers a convenient way to generate multi-soliton solutions.
Let $\{\lambda_j\}_{j=1}^{N}$ be a set of distinct complex spectral parameters of the form
\begin{equation}
\lambda_j = \epsilon_j + i\eta_j, \qquad
\epsilon_j,\eta_j \in \mathbb{R}, \quad
j = 1,2,\ldots,N,
\end{equation}
where we assume $\lambda_i \neq \lambda_j$ for $i \neq j$, meaning that the spectral parameters are distinct complex numbers. This excludes exact spectral degeneracy while still allowing different parameters to share the same real part, corresponding to solitons propagating with identical velocities but different amplitudes.
For each $\lambda_j$, define the phase function
$$
\Theta_j(x,t)
=
\frac{i}{3}
\bigl(\lambda_j x + \lambda_j^2 t\bigr),
\qquad
j=1,2,\ldots,N,
$$
and the corresponding eigenfunction components are chosen in exponential form,
$$
\phi_{j1} = e^{-2\Theta_j},
\qquad
\phi_{j2} = c_{1j} e^{\Theta_j},
\qquad
\phi_{j3} = c_{2j} e^{\Theta_j},
$$
where $c_{1j}$ and $c_{2j}$ are complex constants.
The interaction among the \(N\) eigenfunctions is organized through the \(N \times N\) Gram matrix
\[
M_N=\left(M_{ij}\right)_{1\le i,j\le N},
\]
whose entries are defined through the Hermitian inner products of the eigenfunctions associated with the spectral parameters.
$$
M_{ij}
=
\frac{
\phi_{i1}^{*}\phi_{j1}
+
\phi_{i2}^{*}\phi_{j2}
+
\phi_{i3}^{*}\phi_{j3}
}
{\lambda_j - \lambda_i^{*}},
\qquad
i,j=1,2,\ldots,N.
$$
Since $\operatorname{Im}(\lambda_j)\neq 0$, the denominator does not vanish and $\det(\mathbf{M}_N)$ is well-defined.
The $N$th-order field components are expressed as ratios of bordered Gram-determinants,
\begin{equation}\label{gram-general}
q_1^{(N)}(x,t)
=
\frac{\det\bigl(\mathbf{M}_N^{(1)}\bigr)}
{\det\bigl(\mathbf{M}_N\bigr)},
\qquad
q_2^{(N)}(x,t)
=
\frac{\det\bigl(\mathbf{M}_N^{(2)}\bigr)}
{\det\bigl(\mathbf{M}_N\bigr)},
\end{equation}
here $\mathbf{M}_N^{(1)}$ and $\mathbf{M}_N^{(2)}$ are the $(N+1)\times(N+1)$ bordered matrices defined by
\[
\mathbf{M}_N^{(1)} =
\begin{pmatrix}
\mathbf{M}_N & \mathbf{C}_2^{*} \\
\mathbf{C}_1^{T} & 0
\end{pmatrix},
\qquad
\mathbf{M}_N^{(2)} =
\begin{pmatrix}
\mathbf{M}_N & \mathbf{C}_3^{*} \\
\mathbf{C}_1^{T} & 0
\end{pmatrix},
\]
where
\[
\mathbf{C}_1 =
(\phi_{11},\ldots,\phi_{N1})^{T},
\quad
\mathbf{C}_2 =
(\phi_{12},\ldots,\phi_{N2})^{T},
\quad
\mathbf{C}_3 =
(\phi_{13},\ldots,\phi_{N3})^{T}.
\]
In applications, it is often convenient to select spectral parameters with identical real parts but distinct imaginary parts, for example
$$
\lambda_1=\epsilon_1+i\eta_1,
\quad
\lambda_2=\epsilon_1+i\eta_2,
\quad
\lambda_3=\epsilon_2+i\eta_3,
\quad
\lambda_4=\epsilon_2+i\eta_4,
\ \ldots
$$
under this choice, solitons corresponding to the same real part $\epsilon_j$ propagate with equal velocity $v_j=-2\epsilon_j$, while different imaginary parts determine their amplitudes and internal profiles. Such a configuration is useful for constructing bound-states without invoking spectral degeneracy.
The determinant formulation above remains valid for arbitrary $N$. Although the explicit expanded expressions increase in size as $N$ grows, the matrix structure itself remains unchanged. Therefore, the Gram-determinant representation provides a unified and systematic approach to generating higher-order nondegenerate soliton solutions.

\subsection{Connection Between Iterative DT and Gram-determinants}
The repeated application of one-fold Darboux transformations generates a hierarchy of projector terms whose coefficients are expressed through Hermitian inner products of the associated eigenfunctions. These inner
products form a Gram-type matrix structure, allowing the iterative Darboux representation to be recast into bordered Gram-determinants.
\textit{The above statement illustrates how the
repeated projection-operator structure arising
from the iterative Darboux transformation
naturally evolves into the Gram-determinant
representation.}
\textbf{Theorem 1 (General N-fold Nondegenerate Soliton Solution)}\\
Let $\{\lambda_j\}_{j=1}^{N}$ be a set of distinct complex spectral parameters satisfying
$\operatorname{Im}(\lambda_j)>0$, and let the corresponding eigenfunctions of the
Lax pair be denoted by
\[
\phi_j=(\phi_{j1},\phi_{j2},\phi_{j3})^{T},
\qquad j=1,2,\ldots,N.
\]
The Gram matrix associated with these eigenfunctions is defined as
\[
M_N=(M_{ij})_{1\le i,j\le N},
\]
whose entries are given by
\[
M_{ij}
=
\frac{
\phi_{i1}^{*}\phi_{j1}
+\phi_{i2}^{*}\phi_{j2}
+\phi_{i3}^{*}\phi_{j3}
}
{\lambda_j-\lambda_i^{*}}.
\]
Let $M_N^{(1)}$ and $M_N^{(2)}$ denote the bordered Gram matrices introduced in
Eq.~(34), then the functions
\[
q_1^{(N)}(x,t)
=
\frac{\det(M_N^{(1)})}
{\det(M_N)},
\qquad
q_2^{(N)}(x,t)
=
\frac{\det(M_N^{(2)})}
{\det(M_N)},
\]
constitute an $N$-soliton solution of the Manakov system~\eqref{eq1a}--\eqref{eq1b}.\\
\textbf{Proof.}
The result follows from the $N$-fold Darboux transformation associated with the
Lax pair~\eqref{lax_spatial}--\eqref{lax_temporal}.
Each one-fold Darboux transformation generates a rank-one projector
constructed from an eigenfunction of the associated spectral problem.
Repeated application of the Darboux transformation preserves the
zero-curvature condition
\[
U_t-V_x+[U,V]=0,
\]
and therefore preserves the integrability of the Manakov system.
After $N$ successive iterations, the transformed fields involve
multiple inner products between the eigenfunctions corresponding to
the spectral parameters $\lambda_1,\lambda_2,\ldots,\lambda_N$.
These inner products can be systematically organized into the Gram matrix
$M_N$.
The Gram matrix entries arise naturally from the Hermitian inner products
generated during successive Darboux iterations, while the bordered Gram
determinants represent the numerator corrections associated with the
transformed field variables. Consequently, the repeated projector
structure of the N-fold Darboux transformation can be reorganized into
a compact determinant representation.
Using standard determinant identities, including Jacobi's identity and
Sylvester's determinant theorem, the iterated projector expressions are
rewritten in terms of the bordered Gram-determinants $M_N^{(1)}$ and
$M_N^{(2)}$. These identities establish the necessary relations among the
determinant minors and guarantee consistency with the bilinear structure
generated by the Darboux transformation.
Since the transformed Darboux matrix remains gauge-equivalent to the
original Lax pair, the zero-curvature condition
\[
U_t-V_x+[U,V]=0
\]
is preserved under the N-fold transformation. Therefore, the determinant
expressions obtained from the Gram representation satisfy the same
compatibility condition as the original Manakov system. Hence,
\(q_1^{(N)}\) and \(q_2^{(N)}\) given by Eq.~\eqref{gram-general} constitute a valid
N-soliton solution of the Manakov system.
\subsection{Three-fold DT}
To construct the three-fold nondegenerate solitons, we begin with the same trivial seed solution $q_1=q_2=0$, and write the third-order soliton solution using the general Gram-determinant form given by Eq. \eqref{gram-general} of the following form:
\begin{equation}
q_1^{(3)}(x,t)=\frac{\det(\mathbf{M}_3^{(1)})}{\det(\mathbf{M}_3)}, 
\qquad 
q_2^{(3)}(x,t)=\frac{\det(\mathbf{M}_3^{(2)})}{\det(\mathbf{M}_3)}.
\label{3rd-order solution}
\end{equation}
Since the Darboux transformation is constructed on the trivial seed solution $q_1=q_2=0$, the transformed fields reduce to pure Gram-determinant expressions without an additional background term.
We denote the $3\times3$ Gram matrix associated with the three eigenfunctions for the three-fold DT as $\mathbf{M}_3$, which is of the following form:
\begin{equation}\label{m3}
\mathbf{M}_3=
\begin{pmatrix}
M_{11} & M_{12} & M_{13}\\
M_{21} & M_{22} & M_{23}\\
M_{31} & M_{32} & M_{33}
\end{pmatrix}.
\end{equation}
To construct the explicit third-order nondegenerate solitons, we further developed two augmented Gram-type matrices by appending an additional column and row in the above $\mathbf{M}_3$ matrix given by \eqref{m3}. The superscripts $^{(1)}$ and $^{(2)}$ in $\mathbf{M}_3$ matrices represented in determinants~\eqref{3rd_num_1} and ~\eqref{3rd_num_2} correspond to the numerators of the field components $q_1^{(3)}$ and $q_2^{(3)}$ which show that the augmentation corresponds to the second and third components of the eigenfunctions $\lambda_1$, $\lambda_2$ and $\lambda_3$, denoted by $\phi_{j2}$ and $\phi_{j3}$, where $j=1,2,3$ respectively.
\begin{equation}
\mathbf{M}_3^{(1)}=
\begin{pmatrix}
M_{11} & M_{12} & M_{13} & \phi_{12}^{*}\\
M_{21} & M_{22} & M_{23} & \phi_{22}^{*}\\
M_{31} & M_{32} & M_{33} & \phi_{32}^{*}\\
\phi_{11} & \phi_{21} & \phi_{31} & 0
\end{pmatrix},\label{3rd_num_1}
\end{equation}

\begin{equation}
\mathbf{M}_3^{(2)}=
\begin{pmatrix}
M_{11} & M_{12} & M_{13} & \phi_{13}^{*}\\
M_{21} & M_{22} & M_{23} & \phi_{23}^{*}\\
M_{31} & M_{32} & M_{33} & \phi_{33}^{*}\\
\phi_{11} & \phi_{21} & \phi_{31} & 0
\end{pmatrix}.\label{3rd_num_2}
\end{equation}
The Gram matrix elements $M_{ij}$ $(i,j=1,2,3)$ are defined as
\begin{equation}
\label{Mij}
M_{ij}=
\frac{
\phi_{i1}^{*}\phi_{j1}
+
\phi_{i2}^{*}\phi_{j2}
+
\phi_{i3}^{*}\phi_{j3}
}
{\lambda_j-\lambda_i^{*}},
\end{equation}
 where the explicit expressions are given as
$$
M_{11}=
\frac{
\phi_{11}^{*}\phi_{11}
+
\phi_{12}^{*}\phi_{12}
+
\phi_{13}^{*}\phi_{13}
}
{\lambda_1-\lambda_1^{*}},
\quad
M_{12}=
\frac{
\phi_{11}^{*}\phi_{21}
+
\phi_{12}^{*}\phi_{22}
+
\phi_{13}^{*}\phi_{23}
}
{\lambda_2-\lambda_1^{*}},
\quad
M_{13}=
\frac{
\phi_{11}^{*}\phi_{31}
+
\phi_{12}^{*}\phi_{32}
+
\phi_{13}^{*}\phi_{33}
}
{\lambda_3-\lambda_1^{*}}
$$

$$
M_{21}=
\frac{
\phi_{21}^{*}\phi_{11}
+
\phi_{22}^{*}\phi_{12}
+
\phi_{23}^{*}\phi_{13}
}
{\lambda_1-\lambda_2^{*}},
\quad
M_{22}=
\frac{
\phi_{21}^{*}\phi_{21}
+
\phi_{22}^{*}\phi_{22}
+
\phi_{23}^{*}\phi_{23}
}
{\lambda_2-\lambda_2^{*}},
\quad
M_{23}=
\frac{
\phi_{21}^{*}\phi_{31}
+
\phi_{22}^{*}\phi_{32}
+
\phi_{23}^{*}\phi_{33}
}
{\lambda_3-\lambda_2^{*}}
$$

$$
M_{31}=
\frac{
\phi_{31}^{*}\phi_{11}
+
\phi_{32}^{*}\phi_{12}
+
\phi_{33}^{*}\phi_{13}
}
{\lambda_1-\lambda_3^{*}},
\quad
M_{32}=
\frac{
\phi_{31}^{*}\phi_{21}
+
\phi_{32}^{*}\phi_{22}
+
\phi_{33}^{*}\phi_{23}
}
{\lambda_2-\lambda_3^{*}},
\quad
M_{33}=
\frac{
\phi_{31}^{*}\phi_{31}
+
\phi_{32}^{*}\phi_{32}
+
\phi_{33}^{*}\phi_{33}
}
{\lambda_3-\lambda_3^{*}}
$$
The eigenfunctions corresponding to the spectral parameters
$\lambda_1$, $\lambda_2$, and $\lambda_3$ are given by
\[
\phi_{11}(x,t)=e^{-2\Theta_1(x,t)}, \qquad
\phi_{12}(x,t)=c_{11}e^{\Theta_1(x,t)}, \qquad
\phi_{13}(x,t)=c_{21}e^{\Theta_1(x,t)},
\]
\[
\phi_{21}(x,t)=e^{-2\Theta_2(x,t)}, \qquad
\phi_{22}(x,t)=c_{12}e^{\Theta_2(x,t)}, \qquad
\phi_{23}(x,t)=c_{22}e^{\Theta_2(x,t)},
\]
and
\[
\phi_{31}(x,t)=e^{-2\Theta_3(x,t)}, \qquad
\phi_{32}(x,t)=c_{13}e^{\Theta_3(x,t)}, \qquad
\phi_{33}(x,t)=c_{23}e^{\Theta_3(x,t)}.
\]
The corresponding phase variables take the form
\[
\Theta_1(x,t)=\frac{i}{3}(\lambda_1x+\lambda_1^2t), \qquad
\Theta_2(x,t)=\frac{i}{3}(\lambda_2x+\lambda_2^2t), \qquad
\Theta_3(x,t)=\frac{i}{3}(\lambda_3x+\lambda_3^2t),
\]
where the spectral parameters are chosen as
\[
\lambda_1=\epsilon_1+i\eta_1,\qquad
\lambda_2=\epsilon_1+i\eta_2,\qquad
\lambda_3=\epsilon_2+i\eta_3,
\]
with $\epsilon_1$, $\epsilon_2$, $\eta_1$, $\eta_2$, and $\eta_3$ being real constants.
\subsection{Four-fold DT}
Proceeding similarly, we obtained the fourth-order soliton solutions in the Gram-determinant form
\begin{equation}
q^{(4)}_1(x,t)=\frac{\det\left(M^{(1)}_4\right)}{\det(M_4)}, 
\qquad 
q^{(4)}_2(x,t)=\frac{\det\left(M^{(2)}_4\right)}{\det(M_4)} ,\label{fourth_order_gram}
\end{equation} 
here $M_4$ denotes the $4\times4$ Gram matrix associated with the four eigenfunctions involved in the transformation and is given by
\begin{equation}
M_4=
\begin{pmatrix}
M_{11} & M_{12} & M_{13} & M_{14}\\
M_{21} & M_{22} & M_{23} & M_{24}\\
M_{31} & M_{32} & M_{33} & M_{34}\\
M_{41} & M_{42} & M_{43} & M_{44}
\end{pmatrix},
\end{equation}
and the two augmented Gram-type matrices are given as
\begin{equation}
M^{(1)}_4=
\begin{pmatrix}
M_{11} & M_{12} & M_{13} & M_{14} & \phi_{12}^{*}\\
M_{21} & M_{22} & M_{23} & M_{24} & \phi_{22}^{*}\\
M_{31} & M_{32} & M_{33} & M_{34} & \phi_{32}^{*}\\
M_{41} & M_{42} & M_{43} & M_{44} & \phi_{42}^{*}\\
\phi_{11} & \phi_{21} & \phi_{31} & \phi_{41} & 0
\end{pmatrix},
\end{equation}

\begin{equation}
M^{(2)}_4=
\begin{pmatrix}
M_{11} & M_{12} & M_{13} & M_{14} & \phi_{13}^{*}\\
M_{21} & M_{22} & M_{23} & M_{24} & \phi_{23}^{*}\\
M_{31} & M_{32} & M_{33} & M_{34} & \phi_{33}^{*}\\
M_{41} & M_{42} & M_{43} & M_{44} & \phi_{43}^{*}\\
\phi_{11} & \phi_{21} & \phi_{31} & \phi_{41} & 0
\end{pmatrix}.
\end{equation}
 The corresponding Gram matrix elements $M_{ij}$ $(i,j=1,2,3,4)$ are defined by
$$
M_{ij}=
\frac{\phi_{i1}^{*}\phi_{j1}+\phi_{i2}^{*}\phi_{j2}+\phi_{i3}^{*}\phi_{j3}}
{\lambda_j-\lambda_i^{*}} ,
$$
and the expressions are explicitly shown as
\[
\setlength{\arraycolsep}{8pt}
\begin{array}{ll}
M_{11} = \dfrac{\phi_{11}^{*}\phi_{11}+\phi_{12}^{*}\phi_{12}+\phi_{13}^{*}\phi_{13}}{\lambda_1-\lambda_1^{*}}, &
M_{12} = \dfrac{\phi_{11}^{*}\phi_{21}+\phi_{12}^{*}\phi_{22}+\phi_{13}^{*}\phi_{23}}{\lambda_2-\lambda_1^{*}}, \\[10pt]

M_{13} = \dfrac{\phi_{11}^{*}\phi_{31}+\phi_{12}^{*}\phi_{32}+\phi_{13}^{*}\phi_{33}}{\lambda_3-\lambda_1^{*}}, &
M_{14} = \dfrac{\phi_{11}^{*}\phi_{41}+\phi_{12}^{*}\phi_{42}+\phi_{13}^{*}\phi_{43}}{\lambda_4-\lambda_1^{*}}, \\[10pt]

M_{21} = \dfrac{\phi_{21}^{*}\phi_{11}+\phi_{22}^{*}\phi_{12}+\phi_{23}^{*}\phi_{13}}{\lambda_1-\lambda_2^{*}}, &
M_{22} = \dfrac{\phi_{21}^{*}\phi_{21}+\phi_{22}^{*}\phi_{22}+\phi_{23}^{*}\phi_{23}}{\lambda_2-\lambda_2^{*}}, \\[10pt]

M_{23} = \dfrac{\phi_{21}^{*}\phi_{31}+\phi_{22}^{*}\phi_{32}+\phi_{23}^{*}\phi_{33}}{\lambda_3-\lambda_2^{*}}, &
M_{24} = \dfrac{\phi_{21}^{*}\phi_{41}+\phi_{22}^{*}\phi_{42}+\phi_{23}^{*}\phi_{43}}{\lambda_4-\lambda_2^{*}}, \\[10pt]

M_{31} = \dfrac{\phi_{31}^{*}\phi_{11}+\phi_{32}^{*}\phi_{12}+\phi_{33}^{*}\phi_{13}}{\lambda_1-\lambda_3^{*}}, &
M_{32} = \dfrac{\phi_{31}^{*}\phi_{21}+\phi_{32}^{*}\phi_{22}+\phi_{33}^{*}\phi_{23}}{\lambda_2-\lambda_3^{*}}, \\[10pt]

M_{33} = \dfrac{\phi_{31}^{*}\phi_{31}+\phi_{32}^{*}\phi_{32}+\phi_{33}^{*}\phi_{33}}{\lambda_3-\lambda_3^{*}}, &
M_{34} = \dfrac{\phi_{31}^{*}\phi_{41}+\phi_{32}^{*}\phi_{42}+\phi_{33}^{*}\phi_{43}}{\lambda_4-\lambda_3^{*}}, \\[10pt]

M_{41} = \dfrac{\phi_{41}^{*}\phi_{11}+\phi_{42}^{*}\phi_{12}+\phi_{43}^{*}\phi_{13}}{\lambda_1-\lambda_4^{*}}, &
M_{42} = \dfrac{\phi_{41}^{*}\phi_{21}+\phi_{42}^{*}\phi_{22}+\phi_{43}^{*}\phi_{23}}{\lambda_2-\lambda_4^{*}}, \\[10pt]

M_{43} = \dfrac{\phi_{41}^{*}\phi_{31}+\phi_{42}^{*}\phi_{32}+\phi_{43}^{*}\phi_{33}}{\lambda_3-\lambda_4^{*}}, &
M_{44} = \dfrac{\phi_{41}^{*}\phi_{41}+\phi_{42}^{*}\phi_{42}+\phi_{43}^{*}\phi_{43}}{\lambda_4-\lambda_4^{*}}.
\end{array}
\]
The eigenfunctions corresponding to the spectral parameters $\lambda_1$,
$\lambda_2$, $\lambda_3$, and $\lambda_4$ are given as
\[
\phi_{11}(x,t)=e^{-2\Theta_1(x,t)}, \quad
\phi_{12}(x,t)=c_{11}e^{\Theta_1(x,t)}, \quad
\phi_{13}(x,t)=c_{21}e^{\Theta_1(x,t)},
\]
\[
\phi_{21}(x,t)=e^{-2\Theta_2(x,t)}, \quad
\phi_{22}(x,t)=c_{12}e^{\Theta_2(x,t)}, \quad
\phi_{23}(x,t)=c_{22}e^{\Theta_2(x,t)},
\]
\[
\phi_{31}(x,t)=e^{-2\Theta_3(x,t)}, \quad
\phi_{32}(x,t)=c_{13}e^{\Theta_3(x,t)}, \quad
\phi_{33}(x,t)=c_{23}e^{\Theta_3(x,t)},
\]
and
\[
\phi_{41}(x,t)=e^{-2\Theta_4(x,t)}, \quad
\phi_{42}(x,t)=c_{14}e^{\Theta_4(x,t)}, \quad
\phi_{43}(x,t)=c_{24}e^{\Theta_4(x,t)}.
\]
The phase variables are given as
\[
\Theta_1(x,t)=\frac{i}{3}(\lambda_1x+\lambda_1^2t), \quad
\Theta_2(x,t)=\frac{i}{3}(\lambda_2x+\lambda_2^2t), \quad
\Theta_3(x,t)=\frac{i}{3}(\lambda_3x+\lambda_3^2t), \quad
\Theta_4(x,t)=\frac{i}{3}(\lambda_4x+\lambda_4^2t),
\]
where the  corresponding spectral parameters are chosen as
\[
\lambda_1=\epsilon_1+i\eta_1,\quad
\lambda_2=\epsilon_1+i\eta_2,\quad
\lambda_3=\epsilon_2+i\eta_3,\quad
\lambda_4=\epsilon_2+i\eta_4,
\]
with $\epsilon_1$, $\epsilon_2$, $\eta_1$, $\eta_2$, $\eta_3$, and $\eta_4$ being real constants.

From the explicit three-fold and four-fold constructions given above, a clear systematic matrix pattern can be identified in the determinant formulation of the higher-order solutions. In the three-fold Darboux transformation, the soliton solution is represented through a $3\times3$ Gram matrix $M_3$ together with the corresponding bordered determinant matrices $M_3^{(1)}$ and $M_3^{(2)}$, obtained by appending an additional row and column associated with the eigenfunction components. Similarly, in the four-fold Darboux transformation, the same determinant structure is preserved, while the inclusion of an additional eigenfunction and the spectral parameter naturally extends the Gram matrix from the
$3\times3$ form to the $4\times4$ case together with the corresponding
bordered determinant structures. Thus, each successive Darboux iteration introduces a new spectral parameter
together with its associated eigenfunction components, thereby enlarging the Gram matrix appearing in the denominator and the corresponding augmented bordered matrices systematically appearing in the numerator. Continuing this recursive construction in the same manner, the determinant representation can be generalized systematically to the
$N$th-order case while preserving the same Gram-type matrix
structure.
\subsubsection{General $N^{th}$-Order Nondegenerate Soliton Solution:}

The $N^{th}$-order soliton solutions of the Manakov system are expressed in the Gram-determinant form as
\begin{equation}
q_1^{(N)}(x,t)=\frac{\det(\mathbf{M}_N^{(1)})}{\det(\mathbf{M}_N)}, 
\qquad 
q_2^{(N)}(x,t)=\frac{\det(\mathbf{M}_N^{(2)})}{\det(\mathbf{M}_N)}
\end{equation}
here $\mathbf{M}_N$ denotes the $N\times N$ Gram matrix associated with the $N$ eigenfunctions involved in the transformation and is given by
\begin{equation}
\mathbf{M}_N=
\begin{pmatrix}
M_{11} & M_{12} & \cdots & M_{1N} \\
M_{21} & M_{22} & \cdots & M_{2N} \\
\vdots & \vdots & \ddots & \vdots \\
M_{N1} & M_{N2} & \cdots & M_{NN}
\end{pmatrix},
\end{equation}
 and the augmented matrices for $\mathbf{M}_N^{(1)}$ and $\mathbf{M}_N^{(2)}$ are defined as
\begin{equation}
\mathbf{M}_N^{(1)}=
\begin{pmatrix}
M_{11} & M_{12} & \cdots & M_{1N} & \phi_{12}^{*} \\
M_{21} & M_{22} & \cdots & M_{2N} & \phi_{22}^{*} \\
\vdots & \vdots & \ddots & \vdots & \vdots \\
M_{N1} & M_{N2} & \cdots & M_{NN} & \phi_{N2}^{*} \\
\phi_{11} & \phi_{21} & \cdots & \phi_{N1} & 0
\end{pmatrix},
\end{equation}

\begin{equation}
\mathbf{M}_N^{(2)}=
\begin{pmatrix}
M_{11} & M_{12} & \cdots & M_{1N} & \phi_{13}^{*} \\
M_{21} & M_{22} & \cdots & M_{2N} & \phi_{23}^{*} \\
\vdots & \vdots & \ddots & \vdots & \vdots \\
M_{N1} & M_{N2} & \cdots & M_{NN} & \phi_{N3}^{*} \\
\phi_{11} & \phi_{21} & \cdots & \phi_{N1} & 0
\end{pmatrix}.
\end{equation}
The elements $M_{ij}$ of the Gram matrix are defined by
$$
M_{ij}=
\frac{
\phi_{i1}^{*}\phi_{j1}
+
\phi_{i2}^{*}\phi_{j2}
+
\phi_{i3}^{*}\phi_{j3}
}
{\lambda_j-\lambda_i^{*}},
\qquad
i,j=1,2,\ldots,N,
$$
substituting the explicit eigenfunctions into the above definition, the diagonal elements of the Gram matrix take the form
\begin{small}
$$
M_{11}=
\frac{
e^{-2(\Theta_1+\Theta_1^{*})}
+
|c_{11}|^{2}e^{\Theta_1+\Theta_1^{*}}
+
|c_{21}|^{2}e^{\Theta_1+\Theta_1^{*}}
}
{\lambda_1-\lambda_1^{*}},
\quad
M_{22}=
\frac{
e^{-2(\Theta_2+\Theta_2^{*})}
+
|c_{12}|^{2}e^{\Theta_2+\Theta_2^{*}}
+
|c_{22}|^{2}e^{\Theta_2+\Theta_2^{*}}
}
{\lambda_2-\lambda_2^{*}},
$$
$$
M_{33}=
\frac{
e^{-2(\Theta_3+\Theta_3^{*})}
+
|c_{13}|^{2}e^{\Theta_3+\Theta_3^{*}}
+
|c_{23}|^{2}e^{\Theta_3+\Theta_3^{*}}
}
{\lambda_3-\lambda_3^{*}},
\cdots
M_{NN}=
\frac{
e^{-2(\Theta_N+\Theta_N^{*})}
+
|c_{1N}|^{2}e^{\Theta_N+\Theta_N^{*}}
+
|c_{2N}|^{2}e^{\Theta_N+\Theta_N^{*}}
}
{\lambda_N-\lambda_N^{*}}
$$
\end{small}
and the off-diagonal elements corresponding to the interaction between distinct eigenmodes are given, for $i \neq j$, by
$$
M_{ij}=
\frac{
e^{-2(\Theta_i^{*}+\Theta_j)}
+
c_{1i}^{*}c_{1j}e^{\Theta_i^{*}+\Theta_j}
+
c_{2i}^{*}c_{2j}e^{\Theta_i^{*}+\Theta_j}
}
{\lambda_j-\lambda_i^{*}},
\qquad
i\neq j,
\quad
i,j=1,2,\ldots,N.
$$
The eigenfunctions associated with the spectral parameters $\{\lambda_j\}_{j=1}^{N}$ are given explicitly by
$$
\phi_{j1}(x,t)=e^{-2\Theta_j(x,t)},
\qquad
\phi_{j2}(x,t)=c_{1j}e^{\Theta_j(x,t)},
\qquad
\phi_{j3}(x,t)=c_{2j}e^{\Theta_j(x,t)},
\qquad
j=1,2,\ldots,N,
$$
and their corresponding phase functions are defined as
$$
\Theta_j(x,t)=\frac{i}{3}\left(\lambda_j x+\lambda_j^2 t\right),
\qquad
j=1,2,\ldots,N,
$$
with the spectral parameters are chosen in the form
$$
\lambda_j=\epsilon_j+i\eta_j,
\qquad
j=1,2,\ldots,N,
$$
where $\epsilon_j$ and $\eta_j$ are real constants. Although the explicit algebraic expressions become increasingly complicated with increasing $N$, the above Gram-determinant representation provides a systematic and constructive procedure for generating higher-order soliton solutions.
\section{Bound-state Solitons: Second-order Nondegenerate Solitons}
\vspace{\baselineskip}
%%%%%%%%%%%%%%%%%%%%%%%%%%%%%%%%%%%%%%%%%%%%%%%%%%%%%%%%%%%%%%%%%%%
\subsection{Symmetric and Asymmetric Bound-state Solitons}
\begin{figure}[H]
\centering
\includegraphics[width=0.9\linewidth]{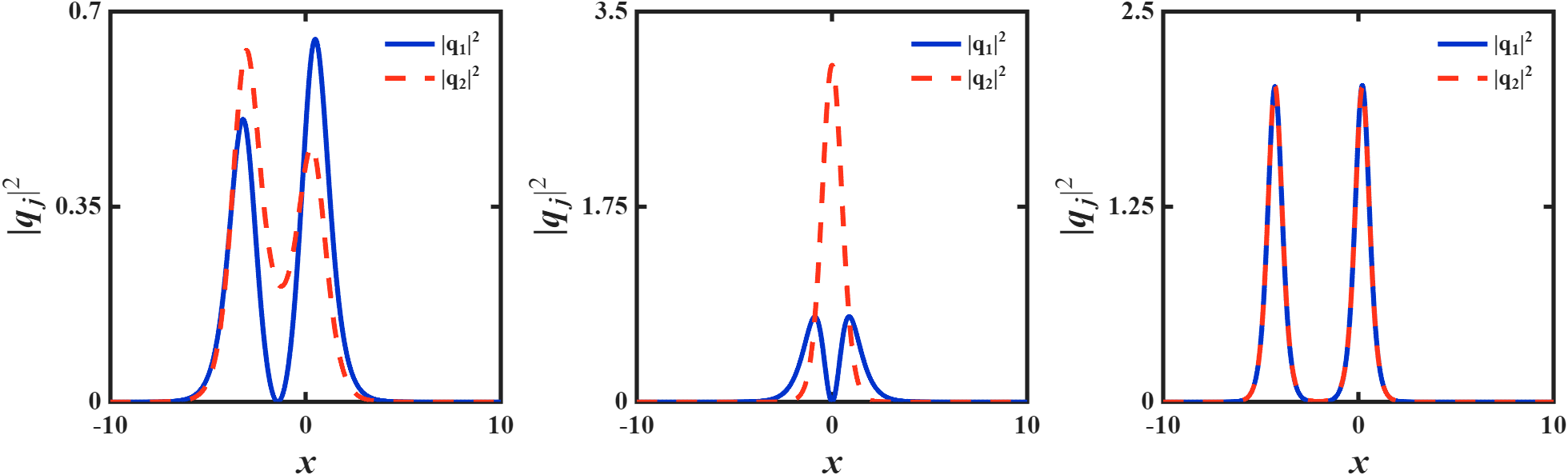}
\caption{The intensity profiles of bound-state solitons
in the two-component coupled system at $t=0$ for different parameter choices.
(a) Asymmetric bound-state structures in both components for 
$\epsilon = 0$, $\eta_1 = 1.01$, $\eta_2 = 1.11$, 
$c_{11} = c_{22} = 1$, and $\delta = -2.8$.
(b) Symmetric configuration where one component exhibits a single-peak
profile and the other a bound-state profile for 
$\epsilon = 0$, $\eta_1 = 1.01$, $\eta_2 = 2.01$, 
$c_{11} = c_{22} = \sqrt{3}/3$, and $\delta = 0$.
(c) Nearly symmetric bound-state profiles in both components for 
$\epsilon = 0$, $\eta_1 = 2.01$, $\eta_2 = 2.011$, 
$c_{11} = c_{22} = 1$, and $\delta = -4.1$.
The solid curve represents $|q_1|^2$ and the dashed curve represents $|q_2|^2$.
}\label{figure-1}
\end{figure}
Figure~\ref{figure-1} illustrates different structural realizations of
bound-state nondegenerate solitons obtained from
Eqs.~\eqref{q1_second order} and \eqref{q2_second order} in the
two-component system at $t=0$. Depending on the parameter selection,
Figure~\ref{figure-1}(a) shows asymmetric bound-state profiles in both
components. In this case, the velocity represented by the real part of the
spectral parameter $\epsilon$ is chosen to be zero, and the amplitude
coefficients controlling the relative intensity are taken as
$c_{11}=c_{22}$, while the imaginary parts of the spectral parameters are
chosen distinctly with $\eta_1 \neq \eta_2$. This difference in the
imaginary parts of the spectral parameters produces interference in the
intensity distribution, resulting in the splitting of a single localized
peak into separated bound-state structures. Here, the distinct but nearly
close values of $\eta_1$ and $\eta_2$ produce asymmetry in
$|q_1(x,t)|^2$ and $|q_2(x,t)|^2$. The coefficients $c_{11}$ and
$c_{22}$ control the relative intensity contribution of the two modes and
thereby influence the peak amplitudes in the two components. Similarly, in
all the cases the amplitude coefficients are chosen to be equal, resulting
in the same relative intensity contribution from both modes. Variations in the values of these coefficients can further modify the relative intensity
distribution between the two components.
Figure~\ref{figure-1}(b) depicts the mixed configuration, where
$|q_1(x,t)|^2$ shows the symmetric bound-state structure, while
$|q_2(x,t)|^2$ exhibits a single-peak, which is primarily due to the
larger difference between the values of $\eta$. This configuration reduces
the interference effect in one component, causing the two localized peaks to merge into a single dominant peak, while the other component retains the bound-state structure. Figure~\ref{figure-1}(c) demonstrates the symmetric bound-state profiles
in both components represented by a blue solid line for $|q_1(x,t)|^2$ and a
dashed red line for $|q_2(x,t)|^2$. Here, the values of $\eta$ are chosen
to be nearly equal. The amplitude coefficients $c_{ij}$ are also taken to
be equal, resulting in nearly symmetric profiles in both components. It can
also be observed that the localization peaks are shifted away from the
origin due to the parameter $\delta$, which controls the relative phase
shift, as seen in Figure~\ref{figure-1}(a) and Figure~\ref{figure-1}(c). In contrast to the degenerate case, where both components share identical
spectral parameters and exhibit symmetric single-peak profiles, the
nondegenerate configuration introduces additional degrees of freedom that
enable the formation of bound-state localized structures.
\subsection{Collisions of Bound-state Solitons}
\begin{figure}[H]
\centering
\includegraphics[width=0.9\linewidth]{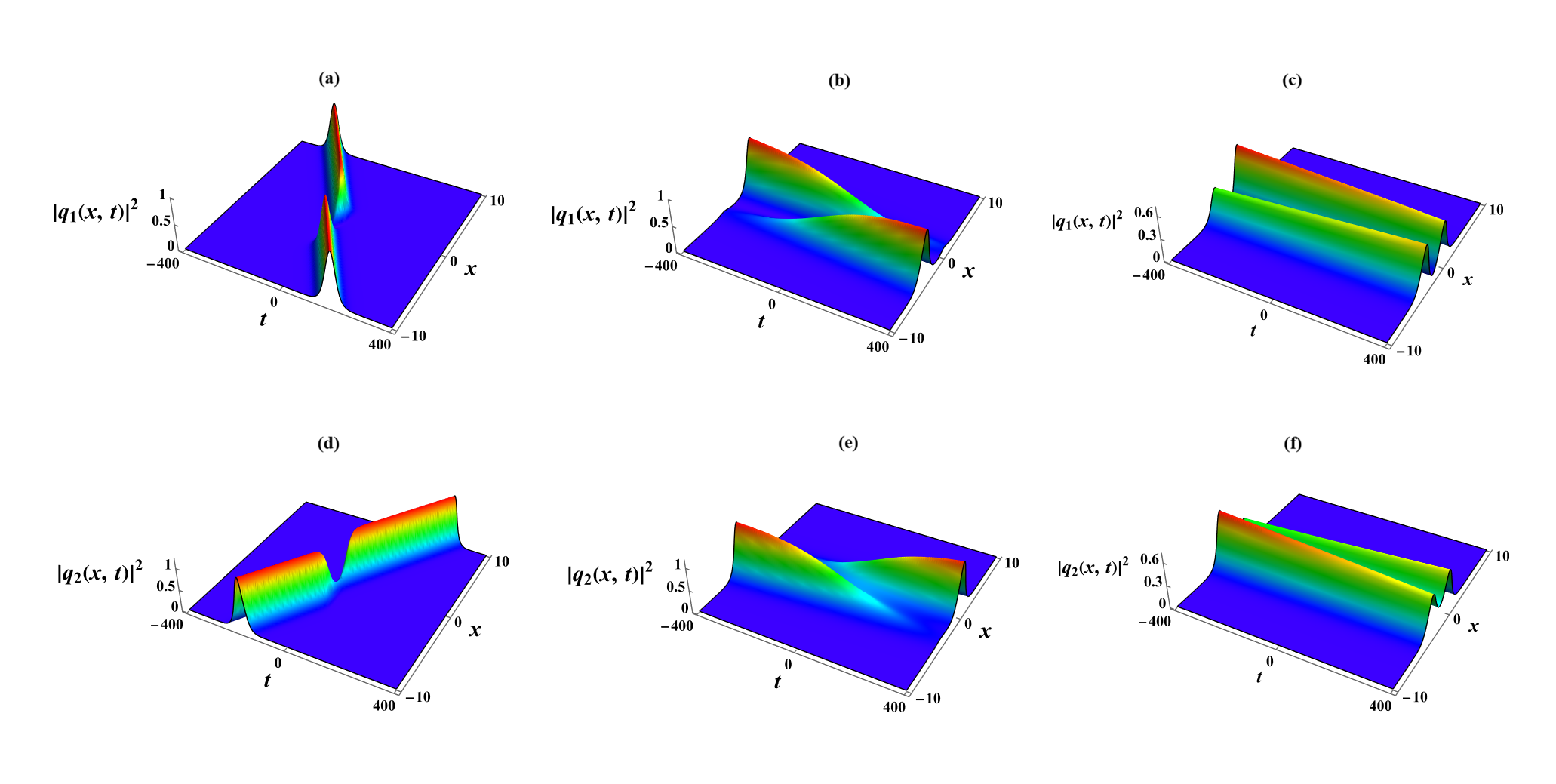}
\caption{Interaction dynamics of nondegenerate two-solitons showing the transition from collision to bound-state formation. 
Panels (a)--(c) show the surface plots of the intensity $|q_1(x,t)|^2$, 
while panels (d)--(f) show $|q_2(x,t)|^2$. 
The parameters are fixed as $\eta_1 = 1$, $\eta_2 = 1.1$, 
$c_{11} = c_{22} = 1$, and $\delta = -2.8$, 
with different relative velocities determined by 
$(\epsilon_1, \epsilon_2)$ as follows: 
(a,d) $\epsilon_1 = 0.02$, $\epsilon_2 = -0.02$; 
(b,e) $\epsilon_1 = 0.001$, $\epsilon_2 = -0.001$; 
(c,f) $\epsilon_1 = 0.0001$, $\epsilon_2 = -0.0001$. 
When the relative velocity between the solitons is reduced, the interaction gradually changes from collisional behaviour to stable nondegenerate bound-state soliton.
} \label{figure-2}
\end{figure}
Figure~\ref{figure-2} shows the evolution of the intensity profiles
illustrating the interaction dynamics of nondegenerate bound-state
solitons for different relative velocities. These behaviours are obtained
by varying the real part of the spectral parameter $\epsilon$, which
determines the soliton velocity through the relation $v_j=-2\epsilon_j$.
The collision dynamics of $|q_1(x,t)|^2$ and $|q_2(x,t)|^2$ are shown in
panels~\ref{figure-2}(a)--\ref{figure-2}(c) and
panels~\ref{figure-2}(d)--\ref{figure-2}(f), respectively.
In Figure~\ref{figure-2}(a) and Figure~\ref{figure-2}(d), the real part of the spectral
parameters $\epsilon$ are chosen with a relatively larger velocity difference. As a
result, the interaction occurs only for a short duration, and the
solitons separate rapidly after collision while retaining their individual structures.
In Figure~\ref{figure-2}(b) and Figure~\ref{figure-2}(e), the difference
between the real parts of the spectral parameters is reduced. Consequently,
the relative velocity decreases, increasing the interaction time between
the two solitons. This produces a broader interaction region and stronger
overlap between the localized structures.
In Figure~\ref{figure-2}(c) and Figure~\ref{figure-2}(f), the real parts
of the spectral parameters are chosen to be nearly equal, such that the
relative velocity approaches nearly zero. In this limit, both the 
solitons propagate with nearly identical velocities and form a persistent
bound-state-like structure characterized by stable propagation.
The top panels correspond to $|q_1(x,t)|^2$, while the bottom panels
represent $|q_2(x,t)|^2$ of the respective bound-state solitons.
The transition from collisional behaviour to bound-state formation is
therefore governed by the relative velocity difference
$|\epsilon_1-\epsilon_2|$. Larger velocity differences produce rapid
collisions with weak interaction, whereas smaller velocity differences
increase the interaction duration and promote a stable bound-state
propagation.
\section{Higher-order Elastic and Inelastic Collisions of Nondegenerate Solitons}
\vspace{\baselineskip}
\subsection{Third-order Elastic Collision of  one Nondegenerate Bound-state Solitons with one Degenerate Soliton}
\begin{figure}[H]
\centering
\includegraphics[width=1\linewidth]{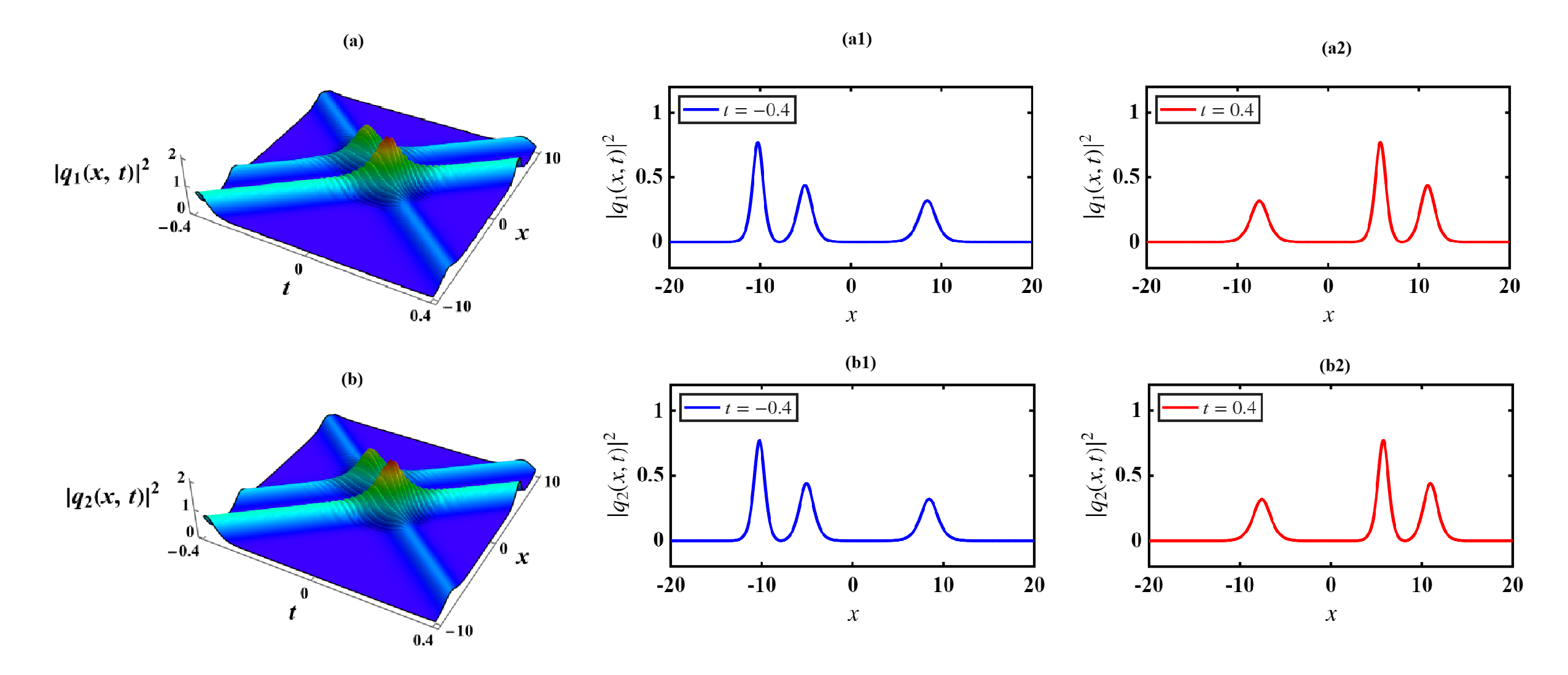}
\caption{Elastic collision between one nondegenerate bound-state solitons 
and a fundamental bright soliton obtained from the third order.
(a) and (b) show the Density evolution of $|q_1(x,t)|^2$ and $|q_2(x,t)|^2$, respectively. (a1) and (b1) depict the Intensity profiles at $t=-0.4$. (a2) and (b2) demonstrate the Intensity profiles at $t=0.4$. The parameters are chosen as 
$\epsilon_1=-10$, $\epsilon_2=10$, 
$\eta_1=1$, $\eta_2=1.1$, $\eta_3=0.8$, 
$c_{11}=1$, $c_{12}=0.5$, $c_{13}=1$, 
$c_{21}=1$, $c_{22}=0.5$, and $c_{23}=1$.
It is seen that the soliton structures are preserved after interaction. \label{3-elastic}}
\end{figure}
Figure \ref{3-elastic} depicts the elastic collision between nondegenerate bound-state solitons and one degenerate soliton obtained by plotting Eq.~\eqref{3rd-order solution} for the choice of parameters mentioned in the figure caption. The real parameters $\epsilon_j$ determine the propagation velocities $v_j=-2\epsilon_j$, while the imaginary parameters $\eta_j$ govern the localization and internal structure of the soliton profiles. For the chosen values $\epsilon_1=-10$ and $\epsilon_2=10$, the two solitons propagate in opposite directions and collide near the central region. Figures~\ref{3-elastic}(a) and \ref{3-elastic}(b) display the dynamical evolution of $|q_1(x,t)|^2$ and $|q_2(x,t)|^2$, respectively. During the interaction process, interference patterns appear in both components due to the coexistence of multiple exponential phase modes associated with different spectral parameters. The three-fold Gram-determinant solution contains several exponential contributions generated from the eigenfunctions corresponding to the spectral parameters $\lambda_1=\epsilon_1+i\eta_1$, $\lambda_2=\epsilon_1+i\eta_2$, and $\lambda_3=\epsilon_2+i\eta_3$. The associated phase functions are defined as
\[
\Theta_j(x,t)=\frac{i}{3}\left(\lambda_jx+\lambda_j^2t\right),
\qquad j=1,2,3,
\]
substituting $\lambda_j=\epsilon_j+i\eta_j$ into the above expression gives
\[
\Theta_j=
\frac{i}{3}\left[(\epsilon_j+i\eta_j)x+(\epsilon_j+i\eta_j)^2t\right],
\]
and expanding the square term gives the expression
\[
(\epsilon_j+i\eta_j)^2
=
\epsilon_j^2-\eta_j^2+2i\epsilon_j\eta_j,
\]
where the following phase function becomes
\[
\Theta_j=
\frac{i}{3}
\left[
\epsilon_jx+(\epsilon_j^2-\eta_j^2)t
\right]
-
\frac{1}{3}
\left[
\eta_jx+2\epsilon_j\eta_jt
\right],
\]
hence
\[
\Theta_j=
-\frac{\eta_j}{3}x
-\frac{2\epsilon_j\eta_j}{3}t
+
\frac{i}{3}
\left[
\epsilon_jx+
(\epsilon_j^2-\eta_j^2)t
\right].
\]
From the above expression, it is clear that the phase function contains two different parts, where the real exponential term
\[
\exp\left[
-\frac{\eta_j}{3}x
-\frac{2\epsilon_j\eta_j}{3}t
\right]
\]
governs the localization and decay of the soliton, while the complex oscillatory phase term
\[
e^{\,i[\epsilon_jx+(\epsilon_j^2-\eta_j^2)t]/3}
\]
controls the wave oscillation during propagation. Since the third-order solution is constructed from several eigenfunctions corresponding to different spectral parameters, the Gram-determinant solution contains multiple exponential phase contributions. During the collision process, these phase modes overlap and interact with each other. Consequently, the solution contains mixed exponential terms associated with different phases. Schematically, these contributions may be represented as
\[
q_j \sim A_1e^{i\Phi_1}+A_2e^{i\Phi_2},
\qquad j=1,2,
\]
where $A_1$ and $A_2$ denote the corresponding amplitude factors and $\Phi_1$, $\Phi_2$ represent the oscillatory phases generated from different spectral parameters.
To understand the origin of the interference pattern, we compute the intensity profile
\[
|q_j|^2=q_jq_j^*,
\]
where the complex conjugate of $q_j$ is
\[
q_j^*\sim A_1^*e^{-i\Phi_1}+A_2^*e^{-i\Phi_2},
\]
now multiplying $q_j$ and $q_j^*$ and expanding the above expression gives
\[
|q_j|^2
=
|A_1|^2+|A_2|^2
+
A_1A_2^*e^{i(\Phi_1-\Phi_2)}
+
A_1^*A_2e^{-i(\Phi_1-\Phi_2)}.
\]
The first two terms correspond to the individual soliton intensities, while the last two terms arise from the interaction between different phase modes. Using Euler's relation
\[
e^{i\theta}+e^{-i\theta}=2\cos\theta,
\]
the mixed terms combine into
\[
2|A_1||A_2|\cos(\Phi_1-\Phi_2),
\]
which generates the oscillatory interference fringes observed in the $(x,t)$-plane.
Using the phase expressions associated with the spectral parameters of the third-order solution, the phase difference becomes
\[
\Phi_1-\Phi_2
=
\frac{1}{3}
\left[
(\epsilon_1-\epsilon_2)x
+
(\epsilon_1^2-\epsilon_2^2+\eta_3^2-\eta_1^2)t
\right].
\]
Therefore, the interference pattern emerges naturally from the cross-phase interaction terms appearing in the modulus square of the higher-order Gram-determinant solution. The overlap between these internal modes during collision produces oscillatory interference structures and transient energy redistribution in the intensity profiles. Further, to confirm the elastic nature of the collision between one nondegenerate bound-state soliton and the degenerate soliton, the corresponding 2D intensity profiles for both components are presented in Figures~\ref{3-elastic}(a1)--(a2) and \ref{3-elastic}(b1)--(b2). The asymptotic analysis presented in Appendix 7.1
shows that the incoming and outgoing coefficient sets
satisfy the elasticity condition~\eqref{eqn_71}--\eqref{eqn_72}.
Therefore the amplitudes and widths of the constituent
localized modes remain unchanged up to phase and
position shifts, confirming the elastic nature of the
interaction.
\subsection{Fourth-order Elastic Collision between two Nondegenerate Bound-state Solitons}
\begin{figure}[H]
\centering
\includegraphics[width=1\linewidth]{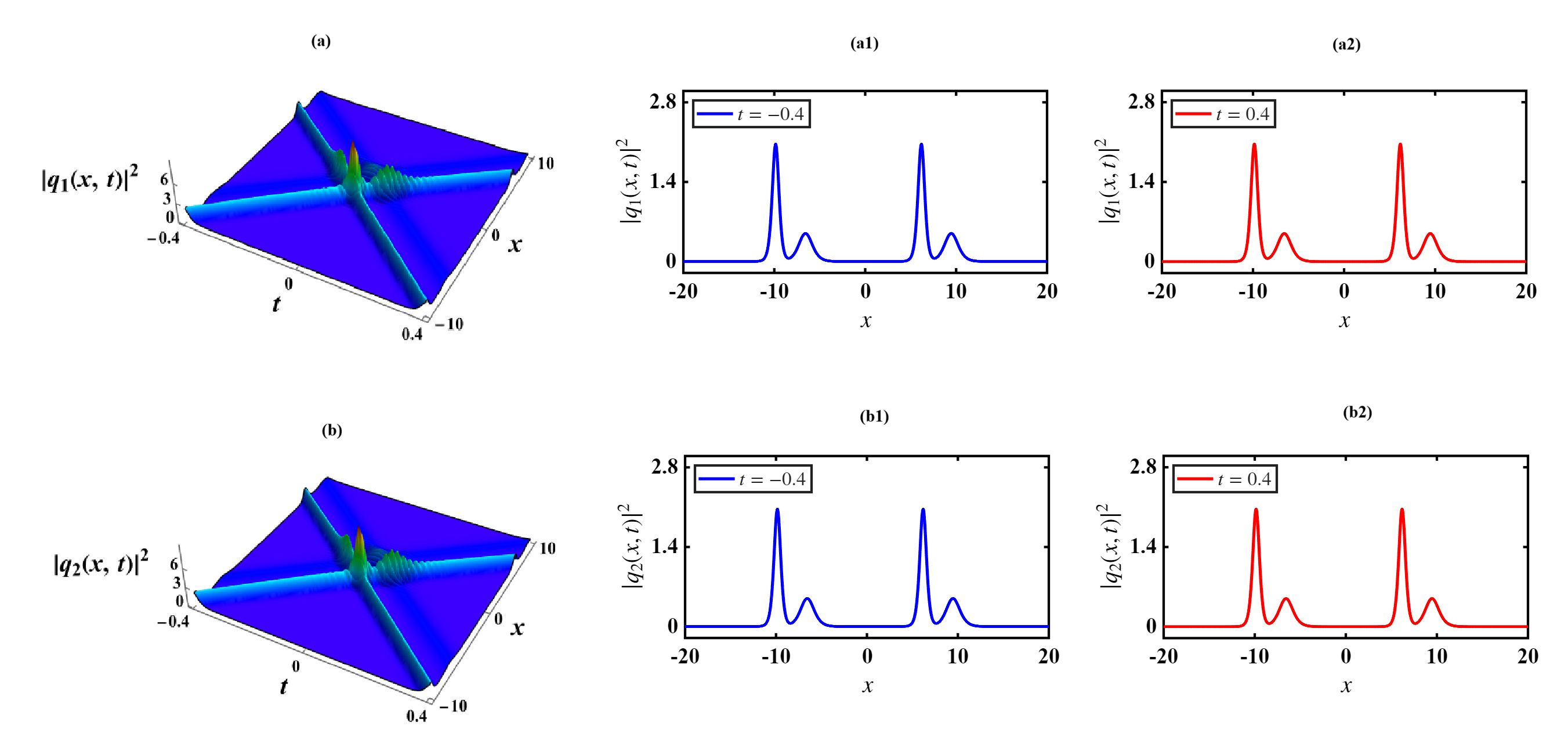}
\caption{Elastic collision between two nondegenerate bound-state solitons derived from the fourth order
(a) and (b) show the Density evolution of $|q_1(x,t)|^2$ and $|q_2(x,t)|^2$, respectively. (a1) and (b1) illustrate the intensity profiles at $t=-0.4$, while (a2) and (b2) depict the Intensity profiles at $t=0.4$. The parameters are chosen as 
$\epsilon_1=10$, $\epsilon_2=-10$, 
$\eta_1=1$, $\eta_2=2$, $\eta_3=1$, $\eta_4=2$, 
$c_{11}=1$, $c_{21}=1$,
$c_{12}=0.05$, $c_{22}=0.05$, 
$c_{13}=1$, $c_{23}=1$,
$c_{14}=0.05$, and $c_{24}=0.05$.
It is seen that the soliton structures are preserved after interaction.
}\label{figure-6}
\end{figure}
In Figure~\ref{figure-6}, we investigate the interaction dynamics of two nondegenerate bound state solitons obtained from the four-fold Darboux transformation given in Eq.~\eqref{fourth_order_gram}. Figures~\ref{figure-6}(a) and \ref{figure-6}(b) display the dynamical evolution of $|q_1(x,t)|^2$ and $|q_2(x,t)|^2$, respectively. For the choice of real parts of the spectral parameters, $\epsilon_1=-10$ and $\epsilon_2=10$, which determine the soliton velocities through the relation $v_j=-2\epsilon_j$, the two nondegenerate bound state solitons propagate in counter-propagating directions and collide near the central interaction region. The imaginary parameters $\eta_j$ govern the localization and internal bound state structure of the soliton profiles.
The amplitude coefficients $c_{ij}$ chosen in the figure caption determine the relative intensity distribution between the two components. In particular, the choice $c_{11}=c_{21}$, $c_{13}=c_{23}$ together with $c_{12}=c_{22}$ and $c_{14}=c_{24}$ ensures that the corresponding interacting modes in both components possess equal relative intensity contributions. As a result, both the modes $|q_1(x,t)|^2$ and $|q_2(x,t)|^2$ exhibit similar interaction behaviour and preserve the symmetric nature of the two bound state soliton configurations throughout the collision process which is confirmed by the 2D subplots shown in Figures~\ref{figure-6}(a1)--(a2) and \ref{figure-6}(b1)--(b2). One can also observe that the peak heights of the two bound-state solitons in both components are asymmetric for the chosen parameter set. However, by suitably tuning the spectral parameters and amplitude coefficients, the relative peak intensities can be tailored to obtain equal-height bound-state structures in both modes, as demonstrated in the following section. Here, one can observe that the chosen spectral parameters satisfy the elastic condition given in~\eqref{eqn_71}--\eqref{eqn_72}, as mentioned in the asymptotic analysis presented in the Appendix of Section~7.1. Hence, the amplitude and width of the respective localized modes remain unchanged after the collision, thereby confirming the elastic nature of the interaction.
\subsection{Controlled Amplitude Modulation of Fourth-order two Nondegenerate Bound-state Soliton Structures}
\begin{figure}[H]
\centering
\includegraphics[width=1\linewidth]{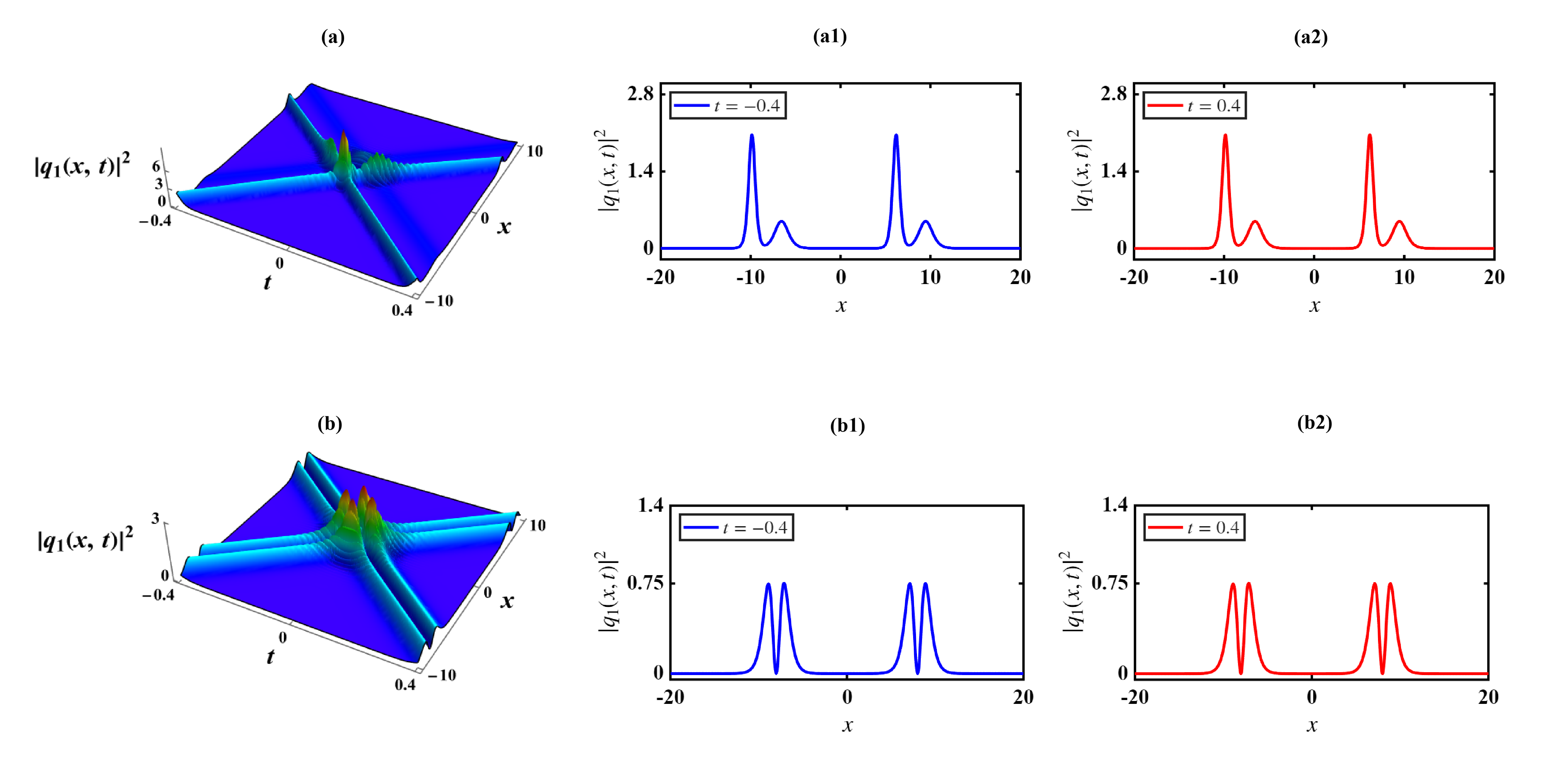}
\caption{Controlled amplitude modulation in the fourth-order two nondegenerate bound-state solitons for $|q_1(x,t)|^2$. Panels (a) and (b) show the three-dimensional
intensity profiles before and after tuning $c_{ij}$, respectively. The corresponding
2D-subplots are presented in (a1)–(a2) and (b1)–(b2) at
$t=-0.4$ and $t=0.4$. Before tuning, the parameters are chosen as
$\epsilon_1=10$, $\epsilon_2=-10$, $\eta_1=1$, $\eta_2=2$, $\eta_3=1$,
$\eta_4=2$, $c_{11}=1$, $c_{21}=1$, $c_{12}=0.05$,
$c_{22}=0.05$, $c_{13}=1$, $c_{23}=1$, $c_{14}=0.05$, and $c_{24}=0.05$.
After tuning the amplitude coefficient $c_{ij}$, which are represented as
$c_{11}=\frac{\sqrt{3}}{3}$, $c_{12}=0$, $c_{13}=\frac{\sqrt{3}}{3}$,
$c_{14}=0$, $c_{21}=0$, $c_{22}=\frac{\sqrt{3}}{3}$, $c_{23}=0$, and
$c_{24}=\frac{\sqrt{3}}{3}$. It can be observed that the peak amplitudes can be tuned in a controlled
manner.
\label{figure-comparison}}
\end{figure}
Figure~\ref{figure-comparison} illustrates the controlled amplitude modulation of the fourth-order nondegenerate bound-state soliton structure. Figures~\ref{figure-comparison}(a) and \ref{figure-comparison}(a1)--(a2) show the soliton profiles before tuning the amplitude coefficients, where the peak heights of the bound-state structures are unequal in both components. In this case, the real parts and imaginary parts of the spectral parameters are kept fixed as shown in the figure caption so that the propagation velocities, localization, and internal bound-state structures remain unchanged. Only the amplitude coefficients $c_{ij}$ are varied in order to control the relative intensity distribution between the interacting modes. Figures~\ref{figure-comparison}(b) and \ref{figure-comparison}(b1)--(b2) depict the soliton profiles after fine-tuning the amplitude coefficients. One can clearly observe that the peak heights of the two bound-state solitons become nearly equal after adjusting the values of $c_{ij}$, while the overall bound-state structure remains preserved. It can also be observed that, even after tuning the amplitude coefficients, rich oscillatory interference structures continue to appear during the interaction process. The origin of these interference patterns arises from the nonlinear coupling between different exponential phase modes associated with distinct spectral parameters, as discussed earlier in Section~5.1. Therefore, the present analysis shows that the amplitude coefficients can be used to tailor the peak-height distribution of the bound-state solitons while preserving the underlying nonlinear interaction dynamics.
\subsection{Fourth-order  Inelastic Collision between two Nondegenerate Bound-state Solitons}
\begin{figure}[H]
\centering
\includegraphics[width=1\linewidth]{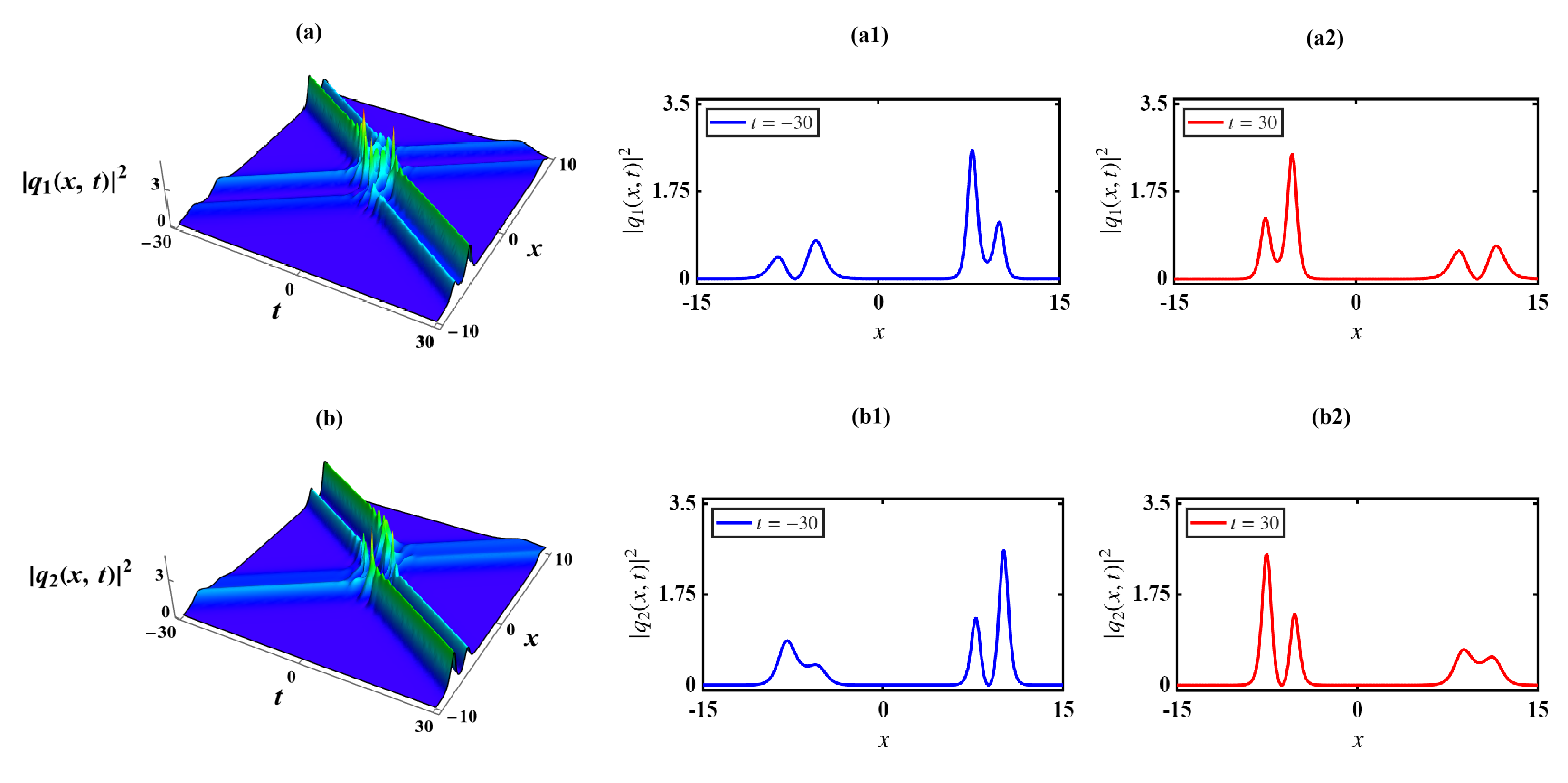}

\caption{Inelastic collision between two nondegenerate bound-state solitons constructed from the fourth-order
(a) and (b) show the Density evolution of $|q_1(x,t)|^2$ and $|q_2(x,t)|^2$, respectively.
(a1) and (b1) depict the Intensity profiles at $t=-30$, whereas
(a2) and (b2) illustrate the Intensity profiles at $t=30$.
The parameters are chosen as 
$\epsilon_1=-\frac{1}{10}$, $\epsilon_2=\frac{1}{10}$,
$\eta_1=1$, $\eta_2=1.2$, $\eta_3=2$, $\eta_4=1.9$,
$c_{11}=1$, $c_{21}=0$, $c_{12}=0$, $c_{22}=1$,
$c_{13}=1$, $c_{23}=0$, $c_{14}=0$, and $c_{24}=1$.
It is seen that the soliton profiles are modified after interaction.
\label{figure-7}}
\end{figure}
\noindent
Figure~\ref{figure-7} illustrates the inelastic interaction between two fourth-order nondegenerate bound-state solitons obtained from the four-fold Darboux transformation. Figures~\ref{figure-7}(a) and \ref{figure-7}(b) display the dynamical evolution of $|q_1(x,t)|^2$ and $|q_2(x,t)|^2$, respectively. The real parts $\epsilon$ of the spectral parameters are chosen with opposite signs, which determine the propagation velocities through the relation $v_j=-2\epsilon_j$. Consequently, the two bound-state solitons propagate in counter-propagating directions and interact near the central collision region. In the present case, the magnitudes of the real parts are chosen to be small that is $\epsilon_1=-0.1$ and $\epsilon_2=0.1$. As a result, the corresponding propagation velocities also become smaller, causing the interaction process to occur over a much longer temporal duration. Therefore, a larger time window that is $t(-30,30)$ is required in order to clearly capture the complete interaction dynamics and post-collision behaviour of the slowly propagating bound-state solitons.
It is well known that the imaginary parameters $\eta_j$ govern the localization widths and internal bound-state structures of the interacting solitons; they are chosen asymmetrically, which influences the interaction dynamics and contributes to the modification of the localized structures after collision.
The amplitude coefficients chosen in the figure caption follow an uneven distribution pattern between the two components. In particular, the coefficients are considered as $c_{11}=c_{22}=c_{13}=c_{24}$, while the remaining set is chosen as $c_{21}=c_{12}=c_{23}=c_{14}$. Due to this unequal distribution of the amplitude coefficients among the interacting internal modes, the relative intensity sharing between the two components becomes asymmetric during interaction. As a result, the post-collision profiles no longer preserve the same amplitudes and internal peak configurations observed before collision.
The intensity distributions before and after interaction are compared through the 2D subplots shown in Figures~\ref{figure-7}(a1)--(a2) and \ref{figure-7}(b1)--(b2).
For the choice of spectral parameters considered here, which generally do not satisfy the elastic condition given in Eqs.~\eqref{eqn_71}--\eqref{eqn_72}, as per the asymptotic analysis, the amplitude and width of the localized modes change after collision, thereby confirming the inelastic nature of the profile.
\subsection{Fifth-order Elastic Collision between two Nondegenerate Bound-state Solitons and one Degenerate Soliton}
\begin{figure}[H]
\centering
\includegraphics[width=1\linewidth]{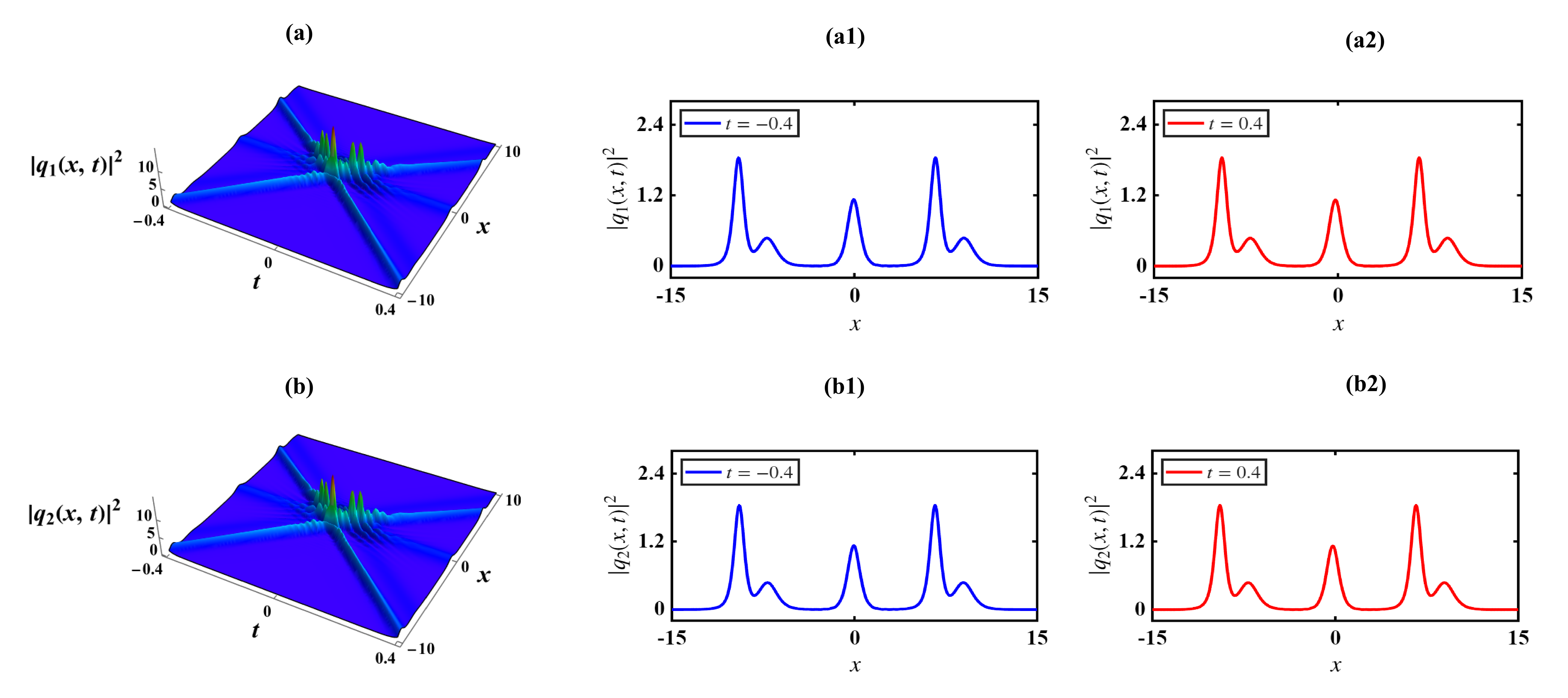}
\caption{
Elastic collision between two nondegenerate bound-state solitons and one degenerate soliton obtained from the fifth order.
(a) and (b) show the Density evolution of $|q_1(x,t)|^2$ and $|q_2(x,t)|^2$, respectively.
(a1) and (b1) Illustrate the Intensity profiles at $t=-0.4$, and
(a2) and (b2) depict the Intensity profiles at $t=0.4$.
The parameters are chosen as 
$\epsilon_1=10$, $\epsilon_2=-10$, $\epsilon_3=0.1$,
$\eta_1=1$, $\eta_2=2$, $\eta_3=1$, $\eta_4=2$, $\eta_5=1.5$,
$c_{11}=\frac{\sqrt{3}}{3}$, $c_{21}=\frac{\sqrt{3}}{3}$,
$c_{12}=0.1$, $c_{22}=0.1$,
$c_{13}=\frac{\sqrt{3}}{3}$, $c_{23}=\frac{\sqrt{3}}{3}$,
$c_{14}=0.1$, $c_{24}=0.1$,
$c_{15}=\frac{\sqrt{3}}{3}$, and $c_{25}=\frac{\sqrt{3}}{3}$.
It is seen that the soliton structures are preserved after interaction.
\label{figure-8}}
\end{figure}
In Figure~\ref{figure-8}, we study the interaction between two nondegenerate 
bound-state solitons and one degenerate soliton generated by the five-fold 
Darboux transformation. For brevity, the explicit form of the fifth-order solution is not presented here. The real parts $\epsilon_1=-10$ and $\epsilon_2=10$, chosen for the two nondegenerate bound-state solitons, cause them to propagate in opposite directions, while the degenerate soliton moves along a vertical trajectory and collides near the central region. Here, the vertical trajectory observed in the density plots corresponds to a degenerate soliton with nearly zero propagation velocity. Since the soliton velocity is governed by the real part of the spectral parameter through \(v=-2\epsilon\), choosing \(\epsilon \approx 0\) leads to negligible spatial drift during evolution, resulting in an almost vertical propagation path in the \((x,t)\) plane. In contrast, the two nondegenerate bound-state solitons exhibit oppositely directed propagation due to the opposite signs of the real parts of the spectral parameters. Also, the amplitude coefficients $c_{ij}$ are chosen following the same pattern as in the previous third- and fourth-order elastic cases. To illustrate the elastic interaction between the two nondegenerate bound-state solitons and the degenerate soliton, the surface plots in Figures~\ref{figure-8}(a) and \ref{figure-8}(b), along with the 2D plots in Figures~\ref{figure-8}(a1)--(a2) and \ref{figure-8}(b1)--(b2), show the space--time evolution of $|q_1(x,t)|^2$ and $|q_2(x,t)|^2$. Despite the complex multi peak structure associated with the fifth-order solution, the localized profiles remain well defined throughout the evolution. Similarly, the chosen set of spectral parameters satisfies the elastic condition as per the asymptotic analysis presented in the appendix. Hence, the localized modes exhibit an elastic nature without undergoing any change in their amplitude and width.
\subsection{Fifth-order Inelastic Collision between two Nondegenerate Bound-state Solitons and one Degenerate Soliton}
\begin{figure}[H]
\centering
\includegraphics[width=1\linewidth]{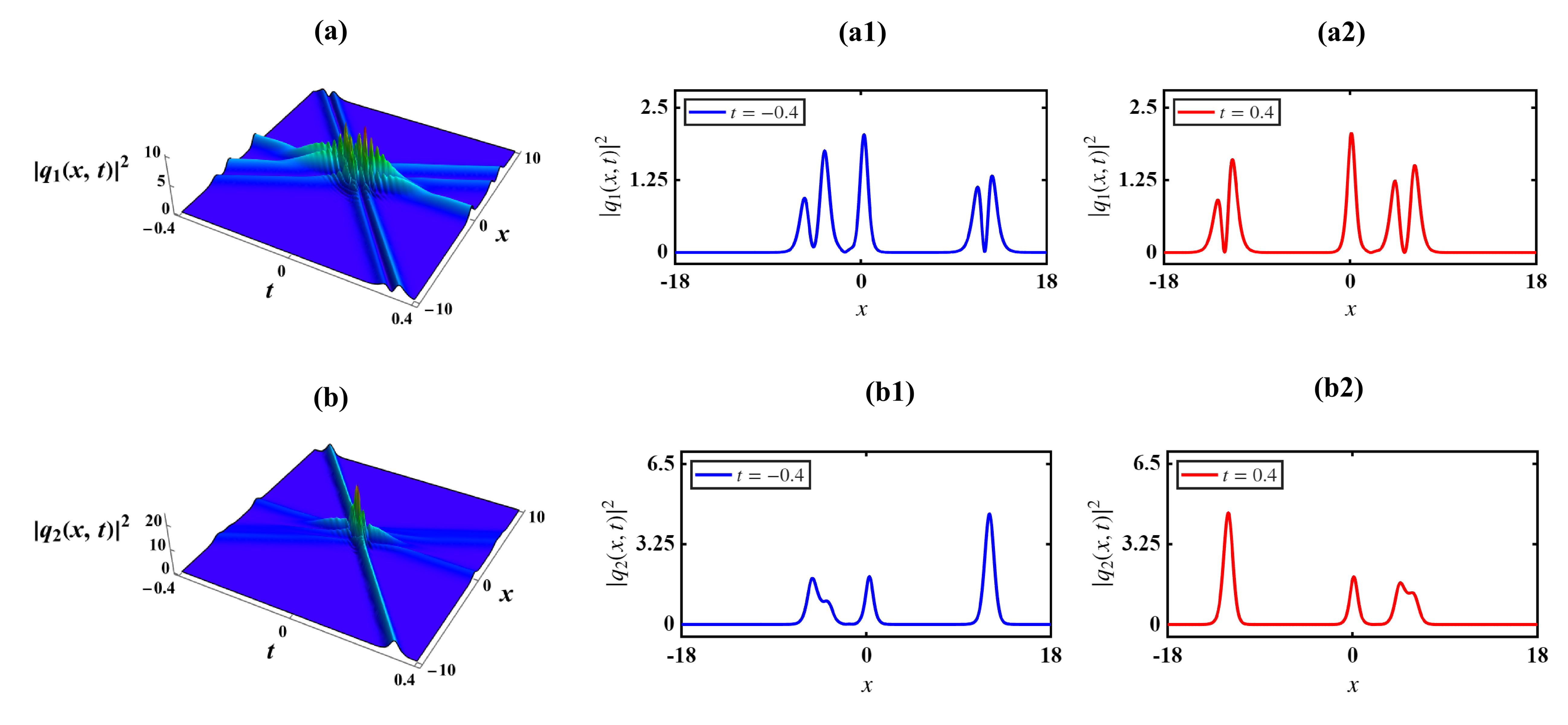}
\caption{
Inelastic interaction of two nondegenerate solitons and one degenerate soliton generated from the fifth-order.
(a) and (b) show the Density distributions of $|q_1(x,t)|^2$ and $|q_2(x,t)|^2$, respectively.
(a1) and (b1) illustrate the intensity Profiles at $t=-0.4$, while
(a2) and (b2) demonstrate the intensity Profiles at $t=0.4$.
The parameters are chosen as 
$\epsilon_1=15$, $\epsilon_2=-6$, $\epsilon_3=0$,
$\eta_1=1.3$, $\eta_2=2.6$, $\eta_3=1.5$, $\eta_4=1.8$, $\eta_5=2$,
$c_{11}=\frac{\sqrt{3}}{3}$, $c_{21}=0.1$,
$c_{12}=0.1$, $c_{22}=\frac{\sqrt{3}}{3}$,
$c_{13}=\frac{\sqrt{3}}{3}$, $c_{23}=0.1$,
$c_{14}=0.1$, $c_{24}=\frac{\sqrt{3}}{3}$,
$c_{15}=1$, and $c_{25}=1$.
After interaction, the internal structures exhibit noticeable modification.\label{figure-9}
}
\end{figure}
In Figure~\ref{figure-9}, we examine the interaction dynamics of two nondegenerate bound-state solitons and one degenerate soliton generated through the five-fold Darboux transformation. Similar to the fourth-order inelastic interaction case, the real parts of the spectral parameters are kept fixed so that the overall propagation characteristics of the interacting solitons remain unchanged. However, the imaginary parts are varied, which modifies the localization widths as well as the internal bound-state structures of the interacting modes. Consequently, the localized peak distributions become asymmetric during propagation.
In addition, the amplitude coefficients $c_{ij}$ are also chosen in unequal manner, following the same procedure adopted in the fourth-order inelastic case. Due to the combined effect of the asymmetric localization properties and unequal intensity distribution among the interacting internal modes, the localized structures no longer preserve their original amplitudes and internal peak configurations after interaction. As a result, the collision exhibits clear inelastic behaviour.
To visualize these interaction dynamics, the surface plots shown in Figures~\ref{figure-9}(a) and \ref{figure-9}(b) illustrate the space--time evolution of $|q_1(x,t)|^2$ and $|q_2(x,t)|^2$, respectively, while the corresponding 2D intensity profiles before and after interaction are presented in Figures~\ref{figure-9}(a1)--(a2) and \ref{figure-9}(b1)--(b2). One can clearly observe that the amplitudes, positions, widths, and internal peaks distributions after collision differ significantly from those before interaction, confirming the inelastic interaction behaviour of the fifth-order nondegenerate bound-state soliton structures. Here also, the chosen set of spectral parameters does not satisfy the elastic condition given in Eqs.~\eqref{eqn_71}--\eqref{eqn_72}, as per the asymptotic analysis. Hence, the soliton profile undergoes visible changes after collision, where the localized modes exhibit variations in their amplitude and width, thereby confirming the inelastic nature of the interaction.

\section{Conclusion}
\vspace{\baselineskip}
The present work establishes a systematic Gram-determinant Darboux transformation framework for constructing higher-order nondegenerate soliton solutions of the integrable Manakov system. The proposed formulation provides a compact and recursive representation for generating $N^{\mathrm{th}}$ order solutions without the rapidly increasing algebraic complexity associated with direct iterative calculations. The determinant structure further highlights how distinct spectral parameters govern the emergence of higher-order localized nonlinear wave patterns and multi peak soliton structures. The obtained higher-order nondegenerate solitons exhibit rich dynamical behaviours such as asymmetric energy redistribution, multi peak localized profiles, velocity-locked propagation, and elastic as well as inelastic collision characteristics. These interaction properties may have potential relevance in nonlinear optical fibres, and related coupled nonlinear wave systems involving controllable energy exchange and multi-mode wave interactions. The present formulation can be extended further to study rogue waves, semi-rational solutions, higher-component coupled systems, nonlocal integrable models, PT-symmetric systems, fractional nonlinear Schr\"odinger-type equations, and spin-orbit-coupled multicomponent systems. The Gram-determinant approach developed here provides a promising framework for investigating stability analysis, perturbation effects, and interaction dynamics of higher-order nonlinear excitations in integrable multi-component systems.
\section{Appendix}
\subsection{Asymptotic Analysis of Two bound-state Soliton Collision}
Although the detailed derivation is presented for the
Four-fold Darboux transformation, the same asymptotic
procedure applies to the third-order and higher-order
Gram-determinant solutions because the determinant
structure differs only in matrix dimension.
Consequently, the elastic and inelastic interaction
criteria derived below remain valid for all higher-order
nondegenerate soliton collisions considered in this work.
The present asymptotic analysis follows the standard multi-soliton scattering framework introduced by Manakov \cite{ref25}.
To investigate the collision dynamics of two nondegenerate bound-state solitons 
obtained through the four-fold Darboux transformation, we perform an asymptotic 
analysis of the solution given by Eq.~\eqref{fourth_order_gram}. The spectral parameters are chosen as 
follows: the first bound-state soliton is generated by the pair $\lambda_1 = 
\epsilon_1 + i\eta_1$ and $\lambda_2 = \epsilon_1 + i\eta_2$, with $c_{21} = 0$, 
$c_{12} = 0$, and $c_{11}$, $c_{22}$ being nonzero complex parameters, while the 
second bound-state soliton is generated by the pair $\lambda_3 = \epsilon_2 + 
i\eta_3$ and $\lambda_4 = \epsilon_2 + i\eta_4$, with $c_{23} = 0$, $c_{14} = 0$, 
and $c_{13}$, $c_{24}$ being nonzero complex parameters. The propagation velocities 
of the two solitons are given by $v_1 = -2\epsilon_1$ and $v_2 = -2\epsilon_2$ 
respectively, and we assume $\epsilon_1 > \epsilon_2$ without loss of generality.

Before proceeding to the asymptotic limits, we first identify the oscillatory 
structure that governs the interaction between the two bound-state solitons. The 
off-diagonal elements of the Gram matrix $M_4$, specifically $M_{13}$, $M_{14}$, 
$M_{23}$, and $M_{24}$, describe the pairwise interaction between the eigenfunctions 
of soliton 1 and soliton 2. These elements contain the cross-phase factors of the form
$e^{\pm i\left[(\epsilon_1 - \epsilon_2)x + 
(\epsilon_1^2 - \epsilon_2^2 + \eta_3^2 - \eta_1^2)t\right]}$
and
$e^{\pm i\left[(\epsilon_1 - \epsilon_2)x + 
(\epsilon_1^2 - \epsilon_2^2 - \eta_2^2 + \eta_4^2)t\right]}$.
Taking the real and imaginary parts of these exponentials, the interaction 
between the two bound-state solitons is governed by the following four 
oscillatory factors:
$\sin\!\left[(\epsilon_1 - \epsilon_2)x + 
(\epsilon_1^2 - \epsilon_2^2 + \eta_3^2 - \eta_1^2)t\right]$,
$\cos\!\left[(\epsilon_1 - \epsilon_2)x + 
(\epsilon_1^2 - \epsilon_2^2 + \eta_3^2 - \eta_1^2)t\right]$,
$\sin\!\left[(\epsilon_1 - \epsilon_2)x + 
(\epsilon_1^2 - \epsilon_2^2 - \eta_2^2 + \eta_4^2)t\right]$,
and
$\cos\!\left[(\epsilon_1 - \epsilon_2)x + 
(\epsilon_1^2 - \epsilon_2^2 - \eta_2^2 + \eta_4^2)t\right]$.
These oscillatory factors reveal that the interaction exhibits multiperiodicity, 
which is a direct consequence of more than two spectral parameters participating 
in the interference process. This stands in sharp contrast to the scalar soliton case, where only a single 
spatial and temporal period governs the interference. The spatial period $D$ 
and the two temporal periods $T_1$ and $T_2$ of the interaction are given 
respectively by $D = \frac{2\pi}{\epsilon_1 - \epsilon_2} = 
\frac{4\pi}{v_1 - v_2}$, $T_1 = \frac{2\pi}{|\epsilon_1^2 - \epsilon_2^2 
+ \eta_3^2 - \eta_1^2|}$, and $T_2 = \frac{2\pi}{|\epsilon_1^2 - 
\epsilon_2^2 - \eta_2^2 + \eta_4^2|}$.
We note that in the special case where $\epsilon_1^2 = 
\epsilon_2^2$, $\eta_1 = \eta_3$, and $\eta_2 = \eta_4$, both temporal 
periods $T_1$ and $T_2$ vanish, and only the spatial interference pattern 
remains visible. Furthermore, as $\epsilon_1 \to \epsilon_2$, the 
spatial period $D \to \infty$, and the two solitons cease to oscillate 
relative to each other, forming instead a stable velocity-locked bound state, 
consistent with the behaviour observed in figure~\ref{figure-2}.
It is worth noting that in the special case where $\epsilon_1^2 = \epsilon_2^2$, 
$\eta_1 = \eta_3$, and $\eta_2 = \eta_4$, both temporal periods $T_1$ and $T_2$ 
vanish, and only the spatial interference pattern remains visible. Furthermore, as 
$\epsilon_1 \to \epsilon_2$, the spatial period $D \to \infty$, and the two 
solitons cease to oscillate relative to each other, forming instead a stable 
velocity-locked bound state, consistent with the behaviour observed in 
figure~\ref{figure-2}.

We now analyze the asymptotic behaviour of the four-fold Darboux transformation 
solution along the trajectory of the first soliton. Along the characteristic line 
$x + 2\epsilon_1 t = \mathrm{const}$, as $t \to +\infty$, we have 
$x + 2\epsilon_2 t \to -\infty$. In this limit, the eigenfunctions associated 
with $\lambda_3$ and $\lambda_4$ simplify as follows:
\begin{equation}
\phi_{31} = e^{-2\Theta_3} \to 0, \qquad 
\phi_{32} = c_{13}e^{\Theta_3} \to 
\begin{pmatrix} 0 \\ c_{13} \\ 0 \end{pmatrix},
\end{equation}
\begin{equation}
\phi_{41} = e^{-2\Theta_4} \to 0, \qquad 
\phi_{43} = c_{24}e^{\Theta_4} \to 
\begin{pmatrix} 0 \\ 0 \\ c_{24} \end{pmatrix}.
\end{equation}
Since the order of iteration of the Darboux transformation matrices can be 
exchanged without affecting the final result, we rewrite the first Darboux matrix 
$T[1]$ constructed from $\lambda_3$ as:
\begin{equation}
T[1] = I - \frac{\lambda_3 - \lambda_3^*}{\lambda - \lambda_3^*}\,P_3.
\end{equation}
Along the line $x + 2\epsilon_1 t = \mathrm{const}$ as $t \to +\infty$, 
this matrix reduces to
\begin{equation}
T[1] \to \mathrm{diag}\!\left(1,\ \frac{\lambda - \lambda_3}{\lambda - \lambda_3^*},
\ 1\right),
\end{equation}
which implies that the transformed eigenfunction corresponding to $\lambda_4$ satisfies
\begin{equation}
\tilde{\phi}_4[1] = T[1]\big|_{\lambda = \lambda_4} 
\begin{pmatrix} 0 \\ 0 \\ c_{24} \end{pmatrix} \to 
\begin{pmatrix} 0 \\ 0 \\ c_{24} \end{pmatrix},
\end{equation}
and consequently yields
\begin{equation}
T[2] \to \mathrm{diag}\!\left(1,\ 1,\ 
\frac{\lambda - \lambda_4}{\lambda - \lambda_4^*}\right).
\end{equation}
Combining the first and second Darboux matrices, we obtain
\begin{equation}
T[2]T[1] \to \mathrm{diag}\!\left(1,\ 
\frac{\lambda - \lambda_3}{\lambda - \lambda_3^*},\ 
\frac{\lambda - \lambda_4}{\lambda - \lambda_4^*}\right),
\end{equation}
from which the transformed eigenfunctions associated with $\lambda_1$ and $\lambda_2$ are obtained as
\begin{equation}
\phi_1[2] = T[2]T[1]\big|_{\lambda=\lambda_1}\phi_1 \to \phi_1^{[+]} = 
\begin{pmatrix} e^{-2\Theta_1} \\ c_{11}^{[+]}e^{\Theta_1} \\ 0 \end{pmatrix},
\end{equation}
\begin{equation}
\phi_2[2] = T[2]T[1]\big|_{\lambda=\lambda_2}\phi_2 \to \phi_2^{[+]} = 
\begin{pmatrix} e^{-2\Theta_2} \\ 0 \\ c_{22}^{[+]}e^{\Theta_2} \end{pmatrix},
\end{equation}
and the phase-shifted coefficients are given by
\begin{equation}
c_{11}^{[+]} = \frac{\lambda_1 - \lambda_3}{\lambda_1 - \lambda_3^*}\,c_{11}, 
\qquad 
c_{22}^{[+]} = \frac{\lambda_2 - \lambda_4}{\lambda_2 - \lambda_4^*}\,c_{22}.
\end{equation}
The four-fold Darboux matrix therefore tends to
\begin{equation}
T_4 \to \left[I - C_2^{[+]}\left(M_2^{[+]}\right)^{-1}
(\lambda I - D_2)^{-1}\left(C_2^{[+]}\right)^\dagger\right] 
\times \mathrm{diag}\!\left(1,\ \frac{\lambda - \lambda_3}{\lambda - \lambda_3^*},\ 
\frac{\lambda - \lambda_4}{\lambda - \lambda_4^*}\right),
\end{equation}
where
\begin{equation}
C_2^{[+]} = \left[\phi_1^{[+]},\ \phi_2^{[+]}\right], \qquad 
D_2 = \mathrm{diag}(\lambda_1^*,\ \lambda_2^*),
\end{equation}
and the reduced Gram matrix is
\begin{equation}
M_2^{[+]} = \begin{pmatrix} 
\dfrac{1 + |c_{11}^{[+]}|^2 e^{6\,\mathrm{Re}(\Theta_1)}}{\lambda_1 - \lambda_1^*} 
& \dfrac{1}{\lambda_2 - \lambda_1^*} \\[12pt] 
\dfrac{1}{\lambda_1 - \lambda_2^*} & 
\dfrac{1 + |c_{22}^{[+]}|^2 e^{6\,\mathrm{Re}(\Theta_2)}}{\lambda_2 - \lambda_2^*} 
\end{pmatrix}.
\end{equation}
Consequently, as $t\to+\infty$ along the line
$x+2\epsilon_1 t=\mathrm{const}$, the first bound-state soliton approaches
\begin{align}
q_1^{[4]} &\to q_1\!\left(x, t;\ \epsilon_1, \eta_1, \eta_2,\ 
c_{11}^{[+]}, c_{22}^{[+]}\right), \\
q_2^{[4]} &\to q_2\!\left(x, t;\ \epsilon_1, \eta_1, \eta_2,\ 
c_{11}^{[+]}, c_{22}^{[+]}\right),
\end{align}
where $q_1$ and $q_2$ on the right-hand side denote the second-order nondegenerate 
bound-state soliton solution given by Eqs.~\eqref{q1_second order}--\eqref{q2_second order}.

Along the same characteristic line $x + 2\epsilon_1 t = \mathrm{const}$, as 
$t \to -\infty$, we have $x + 2\epsilon_2 t \to +\infty$. In this limit, the 
eigenfunctions associated with $\lambda_3$ and $\lambda_4$ behave as:
\begin{equation}
\phi_3 \to \begin{pmatrix} e^{-2\Theta_3} \\ 0 \\ 0 \end{pmatrix}, \qquad 
\phi_4 \to \begin{pmatrix} e^{-2\Theta_4} \\ 0 \\ 0 \end{pmatrix}.
\end{equation}
Following the same procedure as before, the combined Darboux matrix $T[2]T[1]$ 
now reduces to
\begin{equation}
T[2]T[1] \to \mathrm{diag}\!\left(
\frac{(\lambda - \lambda_3)(\lambda - \lambda_4)}
{(\lambda - \lambda_3^*)(\lambda - \lambda_4^*)},\ 1,\ 1\right),
\end{equation}
and the phase-shifted coefficients for soliton 1 in the limit $t \to -\infty$ are 
given as
\begin{equation}
c_{11}^{[-]} = \frac{\lambda_1 - \lambda_3^*}{\lambda_1 - \lambda_3}
\cdot\frac{\lambda_1 - \lambda_4^*}{\lambda_1 - \lambda_4}\,c_{11}, \qquad
c_{22}^{[-]} = \frac{\lambda_2 - \lambda_3^*}{\lambda_2 - \lambda_3}
\cdot\frac{\lambda_2 - \lambda_4^*}{\lambda_2 - \lambda_4}\,c_{22},
\end{equation}
where $t \to -\infty$ along $x + 2\epsilon_1 t = \mathrm{const}$, the 
first bound-state soliton approaches
\begin{align}
q_1^{[4]} &\to q_1\!\left(x, t;\ \epsilon_1, \eta_1, \eta_2,\ 
c_{11}^{[-]}, c_{22}^{[-]}\right), \\
q_2^{[4]} &\to q_2\!\left(x, t;\ \epsilon_1, \eta_1, \eta_2,\ 
c_{11}^{[-]}, c_{22}^{[-]}\right).
\end{align}
By applying the symmetric argument along the characteristic line 
$x + 2\epsilon_2 t = \mathrm{const}$, the asymptotic expressions for the 
second bound-state soliton are obtained in an analogous manner. The phase-shifted 
coefficients for soliton 2 are as follows.
As $t \to +\infty$,
\begin{equation}
c_{13}^{[+]} = \frac{\lambda_3 - \lambda_1^*}{\lambda_3 - \lambda_1}
\cdot\frac{\lambda_3 - \lambda_2^*}{\lambda_3 - \lambda_2}\,c_{13}, \qquad
c_{24}^{[+]} = \frac{\lambda_4 - \lambda_1^*}{\lambda_4 - \lambda_1}
\cdot\frac{\lambda_4 - \lambda_2^*}{\lambda_4 - \lambda_2}\,c_{24},
\end{equation}
when $t \to -\infty$,
\begin{equation}
c_{13}^{[-]} = \frac{\lambda_3 - \lambda_1}{\lambda_3 - \lambda_1^*}\,c_{13}, 
\qquad 
c_{24}^{[-]} = \frac{\lambda_4 - \lambda_2}{\lambda_4 - \lambda_2^*}\,c_{24},
\end{equation}
hence, as $t \to \pm\infty$ along the line 
$x + 2\epsilon_2 t = \mathrm{const}$, the second bound-state soliton approaches
\begin{align}
q_1^{[4]} &\to q_1\!\left(x, t;\ \epsilon_2, \eta_3, \eta_4,\ 
c_{13}^{[\pm]}, c_{24}^{[\pm]}\right), \\
q_2^{[4]} &\to q_2\!\left(x, t;\ \epsilon_2, \eta_3, \eta_4,\ 
c_{13}^{[\pm]}, c_{24}^{[\pm]}\right).
\end{align}
The asymptotic analysis reveals that each bound-state soliton retains its 
functional form before and after the collision, but with modified coefficients 
$c^{[\pm]}$. The collision is elastic if and only if the soliton profiles before 
and after the interaction coincide up to a position shift, which requires
\begin{equation}\label{eqn_71}
|c_{11}^{[+]}| = |c_{11}^{[-]}|\,e^{\eta_1\delta_1}, \qquad 
|c_{22}^{[+]}| = |c_{22}^{[-]}|\,e^{\eta_2\delta_1},
\end{equation}
\begin{equation}\label{eqn_72}
|c_{13}^{[+]}| = |c_{13}^{[-]}|\,e^{\eta_3\delta_2}, \qquad 
|c_{24}^{[+]}| = |c_{24}^{[-]}|\,e^{\eta_4\delta_2},
\end{equation}
and the quantities
\[
\Delta_j
=
\frac{1}{\eta_j}
\ln
\left|
\frac{c_j^{(+)}}{c_j^{(-)}}
\right|,
\]
represent the asymptotic position shifts experienced by the
constituent localized modes during interaction.
When the amplitudes satisfy the elasticity conditions~\eqref{eqn_71}--\eqref{eqn_72}, the collision induces only phase and position
shifts while preserving the overall bound-state profile.
Consequently, the interaction is elastic.
If these conditions are violated, the asymptotic amplitudes
of the constituent modes differ before and after interaction,
leading to energy redistribution between the coupled
components and hence an inelastic collision.
The asymptotic amplitudes before and after collision are
\[
A_j^{-}=|c_j^{-}|,
\qquad
A_j^{+}=|c_j^{+}|,
\]
where the elastic interaction occurs when
\[
\frac{A_j^{+}}{A_j^{-}}
=
e^{\eta_j\delta_j},
\]
which corresponds merely to a translational shift of the
soliton center.
Otherwise
\[
\frac{A_j^{+}}{A_j^{-}}
\neq
e^{\eta_j\delta_j},
\]
and genuine amplitude redistribution occurs,
characterizing an inelastic interaction.
For some real constants $\delta_1$ and $\delta_2$. Since the profile of the 
second-order nondegenerate soliton given by Eqs.~\eqref{q1_second order}--\eqref{q2_second order} is invariant under 
the rescaling $|c_{11}| \to |c_{11}|e^{\eta_1\delta}$ and 
$|c_{22}| \to |c_{22}|e^{\eta_2\delta}$, this condition is the sufficient 
condition for elastic interaction. In the general case, this condition is not 
satisfied, and the collision between the two bound-state solitons is inelastic, 
resulting in energy redistribution between the coupled components, as demonstrated 
in figure~\ref{figure-7}. Under the special parameter conditions where the above 
elasticity condition holds, the collision is elastic and the soliton profiles are 
preserved after interaction, consistent with the behaviour shown in 
figure~\ref{figure-6}.
Therefore, the asymptotic amplitudes, velocities, and bound-state characteristics remain preserved up to phase and position shifts whenever the elasticity conditions are satisfied. Consequently, the interaction is classified as elastic. Violation of these conditions leads to amplitude redistribution among the interacting localized structures and results in an inelastic interaction.

%%%%%%%%%%%%%%%%%%%%%%%%%%%%%%%%%%%%%%%%%%

\ack{
\vspace{\baselineskip}
PSV \& KNS wishes to express their gratitude to the Management of PSG College of Arts and Science for their moral support and encouragement throughout the tenure of this project.}

\roles{
\vspace{\baselineskip}
\noindent K N~Santhiya\\
\noindent Methodology, Software, Visualization, Formal Analysis. \\
\vspace{\baselineskip}
\noindent P S Vinayagam\\
\noindent Conceptualization, Investigation, Methodology, Visualization, Formal Analysis, Writing - Original Draft Preparation, Writing - Review \& Editing, Validation, Resources, Supervision \\
}

\data{
\vspace{\baselineskip}
The data used to describe the article's content is available within the article.}


\begin{thebibliography}{99}
\vspace{\baselineskip}
\bibitem{ref1} Ablowitz M J and Segur H 1981 \textit{Solitons and the Inverse Scattering Transform} (Philadelphia, PA: SIAM)

\bibitem{ref2} Ablowitz M J 2011 \textit{Nonlinear Dispersive Waves: Asymptotic Analysis and Solitons} (Cambridge: Cambridge University Press)

\bibitem{ref3} Ablowitz M J and Musslimani Z H 2021 Integrable nonlinear Schrödinger systems and their soliton dynamics \textit{Phys. Rev.} E \textbf{103} 012214

\bibitem{ref4} Zakharov V E and Gelash A A 2022 Nonlinear stage of modulation instability and soliton interactions \textit{Phys. Rev. Lett.} \textbf{128} 054101

\bibitem{ref5} Derzho O 2019 Waves of complicated structure in nonlinear dispersive media \textit{J. Phys.: Conf. Ser.} \textbf{1268} 012020

\bibitem{ref6} Xia R, Li Y, Tang X and Xu G 2023 Coupling dynamics of dissipative localized structures \textit{Opt. Commun.} \textbf{546} 129996

\bibitem{ref7} Aranson I S and Kramer L 2002 The world of the complex Ginzburg--Landau equation \textit{Rev. Mod. Phys.} \textbf{74} 99

\bibitem{ref8} Trillo S and Wabnitz S 1991 Dynamics of nonlinear modulational instability \textit{Opt. Lett.} \textbf{16} 986

\bibitem{ref9} Zhang H Q 2013 Energy-exchange collisions of vector solitons \textit{Opt. Commun.} \textbf{291} 537

\bibitem{ref10} Radha R, Sakthivinayagam P, 
Shin H J and Porsezian K 2014
Spatiotemporal binary interaction and designer quasi-particle condensates
\textit{Chinese Phys. B}
\textbf{23} 034214

\bibitem{ref11} Karjanto N 2022 Bright soliton solution of nonlinear Schrödinger equation \textit{Mathematics} \textbf{10} 4559

\bibitem{ref12} Al Khawaja U and Bahlouli H 2019 Integrability conditions and solitonic solutions \textit{Commun. Nonlinear Sci. Numer. Simul.} \textbf{69} 248

\bibitem{ref13} Lou Y, Zhang Y, Ye R and Li M 2021 Solitons in nonlocal coupled nonlinear Schrödinger equation \textit{Appl. Math. Comput.} \textbf{409} 126417

\bibitem{ref14} Wang X B and Han B 2021 Solitons in nonlinear systems with higher-order effects \textit{Appl. Math. Lett.} \textbf{113} 106656

\bibitem{ref15} Bibi I, Muhammad S, Shakeel M and Ceesay B 2025 Optical soliton structures and modulation stability \textit{Adv. Math. Phys.} \textbf{2025} 9185387

\bibitem{ref16} Kudryashov N A 2021 Optical solitons of resonant nonlinear Schrödinger equation \textit{Optik} \textbf{244} 166626

\bibitem{ref17} Omar F M, Murad M A S, Mahmood S S, Malik S and Radwan T 2025 Optical solitons and chaos \textit{Sci. Rep.} \textbf{15} 20073

\bibitem{ref18} Liu J Z, Zhang H, Bibi S and Duan W S 2025 High-frequency nonlinear electromagnetic waves \textit{Phys. Plasmas} \textbf{32} 052107

\bibitem{ref19} Chian A C-L, Borotto F A, Hada T, Miranda R A and Munoz P R 2022 Plasma turbulence \textit{Rev. Mod. Plasma Phys.} \textbf{6} 34

\bibitem{ref20} Liu J Z, Zhang H, Han J F and Duan W S 2026 Dark envelope electromagnetic waves \textit{Chaos Solitons Fractals} \textbf{181} 118159

\bibitem{ref21} Peregrine D H 1983 Water waves and nonlinear Schrödinger equations \textit{J. Austral. Math. Soc. Ser. B} \textbf{25} 16

\bibitem{ref22} Maiden M D, Anderson D V, Franco N A, El G A and Hoefer M A 2018 Solitonic dispersive hydrodynamics \textit{Phys. Rev. Lett.} \textbf{120} 144101

\bibitem{ref23} Stepanyants Y A 2019 Internal envelope solitons \textit{Fluids} \textbf{4} 56
\bibitem{ref24} Zakharov V E and Shabat A B 1972
\textit{Exact theory of two-dimensional self-focusing and one-dimensional self-modulation of waves in nonlinear media}
\textit{Sov. Phys. JETP} \textbf{34} 62--69

\bibitem{ref25} Manakov S V 1974 \textit{On the theory of two-dimensional stationary self-focusing of electromagnetic waves} \textit{Sov. Phys.--JETP} \textbf{38} 248--253

\bibitem{ref26} Agrawal G P 2019 \textit{Nonlinear Fiber Optics} 6th edn (San Diego, CA: Academic Press)

\bibitem{ref27} Zhang G, Ling L and Yan Z 2021 Higher-order vector Peregrine solitons \textit{J. Nonlinear Sci.} \textbf{31} 81

\bibitem{ref28} Zhang G, Huang P, Feng B F and Wu C 2023 Rogue waves in vector nonlinear Schrödinger equation \textit{J. Nonlinear Sci.} \textbf{33} 116

\bibitem{ref29} Prinari B 2023 Inverse scattering transform \textit{J. Nonlinear Math. Phys.} \textbf{30} 317

\bibitem{ref30} Yang J 2010 Nonlinear waves \textit{SIAM Rev.} \textbf{52} 429

\bibitem{ref31} Yang J 2017 N-solitons in nonlocal nonlinear Schrödinger equations \textit{Stud. Appl. Math.} \textbf{138} 1

\bibitem{ref32} Akhmediev N N and Ankiewicz A 2002 Solitons around us \textit{Nonlinear Physics} (Berlin: Springer) pp 105--126

\bibitem{ref33} Malomed B A 2005 bound-states of envelope solitons \textit{Prog. Opt.} \textbf{43} 71

\bibitem{ref34} Steiglitz K 2001 Time-gated Manakov solitons \textit{Phys. Rev.} E \textbf{63} 016608

\bibitem{ref35} Leuthold J, Koos C and Freude W 2010 Nonlinear silicon photonics \textit{Nat. Photonics} \textbf{4} 535

\bibitem{ref36} Mollenauer L F and Gordon J P 2006 Solitons in optical fibers \textit{Proc. IEEE} \textbf{94} 131

\bibitem{ref37} Hasegawa A and Matsumoto M 2003 \textit{Optical Solitons in Fibers} (Berlin: Springer)

\bibitem{ref38} Kanna T and Lakshmanan M 2003 Exact soliton solutions, shape changing collisions, and autonomous multisoliton complexes in coupled nonlinear Schrödinger equations \textit{Phys. Rev.} E \textbf{67} 046617

\bibitem{ref39} Pan L, Wang L, Liu L, Sun W and Ren X 2024 Nondegenerate localized waves in the general coupled nonlinear Schrödinger equations \textit{Nonlinearity} \textbf{37} 105016

\bibitem{ref40} Stalin S, Ramakrishnan R and Lakshmanan M 2021 Nondegenerate bright solitons in coupled nonlinear Schrödinger equations \textit{Photonics} \textbf{8} 258

\bibitem{ref41} Yang J 2019 General N-solitons and their dynamics in several nonlocal nonlinear Schrödinger equations \textit{Phys. Lett.} A \textbf{383} 328

\bibitem{ref42} Song N, Lei Y and Cao D 2022 Higher-order soliton dynamics in a generalized nonlinear Schrödinger equation \textit{Acta Mech. Sin.} \textbf{38} 121350

\bibitem{ref43} Baronio F, Degasperis A, Conforti M and Wabnitz S 2012 Solutions of the vector nonlinear Schrödinger equations: evidence for deterministic rogue waves \textit{Phys. Rev. Lett.} \textbf{109} 044102

\bibitem{ref44} Lashkin V M 2021 Fokas--Lenells solitons, breathers and rogue waves on a background of periodic waves \textit{Phys. Rev.} E \textbf{103} 042203

\bibitem{ref45} Porsezian K 1998 Bilinearization of coupled NLS type equations \textit{J. Nonlinear Math. Phys.} \textbf{5} 126

\bibitem{ref46} Radha R, Vinayagam P S 
and Porsezian K 2016
Manipulation of light in a generalized coupled nonlinear Schrödinger equation
\textit{Commun. Nonlinear Sci. Numer. Simul.}
\textbf{39} 87--102
\bibitem{ref47} Vakhnenko O O 1999
\textit{Nonlinear beating excitations on ladder lattice}
\textit{J. Phys. A: Math. Gen.} \textbf{32} 5735
\bibitem{ref48} Vakhnenko O O, Vakhnenko V O and Verchenko A P 2025
\textit{Physical insight into the semi-discrete nonlinear integrable systems with the true and false multicomponentness}
\textit{Chaos Solitons Fractals} \textbf{200} 117043

\bibitem{ref49} Ling L, Zhao L and Guo B 2012 Higher-order rogue waves and external-field-controlled solutions in the coupled nonlinear Schrödinger equations \textit{Phys. Rev.} E \textbf{86} 066606

\bibitem{ref50} Stalin S, Ramakrishnan R and Lakshmanan M 2019 Nondegenerate solitons in coupled nonlinear Schrödinger equations \textit{Phys. Rev. Lett.} \textbf{122} 043901

\bibitem{ref51} Qin Y H, Zhao L C and Ling L 2019 Bound-state solitons of the coupled nonlinear Schrödinger equations \textit{Phys. Rev.} E \textbf{100} 022212

\bibitem{ref52} Stalin S, Ramakrishnan R and Lakshmanan M 2020 Nondegenerate soliton solutions in a generic (N+1)-component coupled nonlinear Schrödinger equation \textit{Phys. Lett.} A \textbf{384} 126201

\bibitem{ref53} Ramakrishnan R, Stalin S and Lakshmanan M 2021 Multihumped (nondegenerate) solitons in a (2+1)-dimensional nonlocal nonlinear Schrödinger equation \textit{J. Phys. A: Math. Theor.} \textbf{54} 14LT01

\bibitem{ref54} An L, Ling L and Zhang X 2023 Solitons and breathers of the fractional Hirota equation \textit{Phys. Lett.} A \textbf{460} 128629

\bibitem{ref55} Cai Y J, Wu J W and Lin J 2022 Nondegenerate N-soliton solutions and their collision dynamics in the vector nonlinear Schrödinger equation \textit{Chaos Solitons Fractals} \textbf{164} 112678
\bibitem{ref56} Matveev V B and Salle M A 1991
\textit{Darboux Transformations and Solitons}
(Berlin: Springer-Verlag)
\bibitem{ref57} Gu C, Hu H and Zhou Z 2005
\textit{Darboux Transformations in Integrable Systems: Theory and their Applications to Geometry}
(Dordrecht: Springer)
\bibitem{ref58} Zhang H Q, Li J, Xu T, Zhang Y X, Hu W and Tian B 2007
\textit{Optical soliton solutions for two coupled nonlinear Schrödinger systems via Darboux transformation}
\textit{Phys. Scr.} \textbf{76} 452--460
\bibitem{ref59} He J, Zhang L, Cheng Y and Li Y 2006
\textit{Determinant representation of Darboux transformation for the AKNS system}
\textit{Sci. China Ser. A Math.} \textbf{49} 1867--1878
\bibitem{ref60} B. Guo, L. Ling and Q. P. Liu, ``Nonlinear Schrödinger equation: Generalized Darboux transformation and rogue wave solutions,'' \textit{Phys. Rev. E} \textbf{85}, 026607 (2012).

\end{thebibliography}
\end{document}